%% file: main.tex
\documentclass[11pt,a4paper]{article}
\pdfoutput=1
\usepackage{jheppub}

\usepackage{float}
\usepackage{afterpage}
\usepackage{amssymb}
\usepackage{amsmath}
\usepackage{multirow}
\usepackage{url}
\usepackage{xcolor}
\usepackage{float}
\usepackage{afterpage}
\usepackage{url}
\usepackage{booktabs}
\usepackage{tikz}
\usetikzlibrary{arrows}
\usepackage{tikz-3dplot}
\usepackage{enumitem}
\usepackage{tabularx}
\usepackage[off]{auto-pst-pdf}
\usepackage{simpler-wick}
\graphicspath{{./figs/}}

\newcommand{\lp}{\left(}
\newcommand{\rp}{\right)}
\newcommand{\bare}{{(0)}}

\numberwithin{equation}{section}
\numberwithin{figure}{section}
\numberwithin{table}{section}

\newcolumntype{C}[1]{>{\centering\arraybackslash}p{#1}}

\title{\boldmath Neural-network analysis of Parton Distribution Functions from Ioffe-time pseudodistributions}

\author[a]{Luigi Del Debbio,\note{Corresponding author.}}
\author[a,1]{Tommaso Giani,}
\author[b]{Joseph Karpie,}
\author[c,d]{Kostas Orginos,} 
\author[c,e]{Anatoly Radyushkin}
\author[f]{and Savvas Zafeiropoulos}

\affiliation[a]{The Higgs Centre for Theoretical Physics, The University of Edinburgh,\\
  Peter Guthrie Tait Road, Edinburgh EH9 3FD, United Kingdom}
\affiliation[b]{Physics Department, Columbia University, New  York City, New York 10027, USA}
\affiliation[c]{Thomas Jefferson National Accelerator Facility, Newport News, Virginia 23606, USA}
\affiliation[d]{Physics Department, College of William and Mary, Williamsburg, Virginia 23187, USA}
\affiliation[e]{Physics Department, Old Dominion University, Norfolk, Virginia 23529, USA}
\affiliation[f]{Aix Marseille Univ, Université de Toulon, CNRS, CPT, Marseille, France}

\emailAdd{luigi.del.debbio@ed.ac.uk}
\emailAdd{tommaso.giani@ed.ac.uk}
\emailAdd{jmk2289@columbia.edu}
\emailAdd{kostas@wm.edu}
\emailAdd{radyush@jlab.org}
\emailAdd{savvas.zafeiropoulos@cpt.univ-mrs.fr}

\abstract{
\noindent We extract two nonsinglet nucleon Parton Distribution Functions  from
lattice QCD data for reduced Ioffe-time pseudodistributions. We
perform such analysis within the {\tt NNPDF} framework, considering data coming
from different lattice ensembles and discussing in detail the treatment of the
different source of systematics involved in the fit. We introduce a recipe for taking care
of systematics and use it to perform our extraction of light-cone PDFs.}

\begin{document}
\maketitle
\flushbottom

\input{sect1}

\input{sect2}

\input{sect3}

\input{sect4}

\input{sect5}
\appendix
\input{appendix}

\input{appendix2}

\bibliographystyle{JHEP}
\bibliography{main}

\end{document}

%% file: sect1.tex
\section{Introduction}
\label{sec:intro}

Parton Distribution Functions (PDFs), encoding the structure of the proton in
terms of quarks and gluons, are one of the main ingredients required to do
precise high-energy phenomenology.  
The available PDFs sets are extracted through global fits over experimental
data~\cite{Ball:2017nwa, Dulat:2015mca, Alekhin:2017kpj, Martin:2009iq,
Buckley:2014ana}. Their non-perturbative nature makes them a natural candidate
for a lattice QCD investigation, however it has been known for a long time that
it is not possible to obtain them directly from first principle computations, due to the Euclidean metric of the lattice. In the last
few years, several methods have been formulated~\cite{Lin:2020rut, Constantinou:2020pek}, which
would allow us to compute on the lattice specific quantities that, in turn, can be
related to PDFs through a factorization theorem. For a detailed discussion of the
theoretical background, we refer the reader to recent reviews such as
Refs.~\cite{Radyushkin:2019mye,Cichy:2018mum, Monahan:2018euv,Ji:2020ect}.

Examples of such quantities are the equal time correlators underlying the
definition of quasi- and pseudo-PDFs \cite{PhysRevLett.110.262002,
Radyushkin:2017cyf}, given by 
\begin{align}
	\label{eq:Ioffe}
	M^\bare_{\mu}\left(z,P\right) &= \langle P |\bar{\psi}^\bare\lp z\rp \gamma_{\mu} \,   
    U^\bare\lp z,0\rp \psi^\bare\lp 0\rp |P\rangle\, ,
\end{align}
with $P$ denoting the momentum of the external proton states, while the suffix
$(0)$ reminds us that these are bare quantities.  The matrix element of the
vector bilocal operator of Eq.~\eqref{eq:Ioffe} can be decomposed  
in terms of two form factors which only depend on the Lorentz invariants $z^2$ and $\nu \equiv - z\cdot P$ as 
\begin{align}
	\label{eq:Ioffedec}	
	M^\bare_{\mu}\left(z,P\right)    
     = 2 P_\mu\,\mathcal{M}^{(0)} \lp \nu,z^2 \rp
    + z_\mu\, \mathcal{N}^{(0)}\lp \nu,z^2 \rp \, .
\end{align}
As pointed out in~\cite{Musch:2010ka}, only the first form factor, $\mathcal{M}^{(0)}$, contains leading twist information. 
This can be seen by choosing a light-cone separation $z=\left(0,z^-,0_{\perp}\right)$ together with
$\gamma^{\mu}=\gamma^{+}$ and $P=\left(P^+,0,0_{\perp}\right)$, then we get 
\begin{align}
	\label{eq:barePDF}
	M^\bare_{+}\left(z, P\right)
	= 2 P_+\,\mathcal{M}^{(0)}\lp \nu,0 \rp
	= 2 P_+\, \int_{-1}^{1} dx\, e^{i x\nu} f^{(0)}\left(x\right)
\end{align}  
with $f^{(0)}\left(x\right)$ being the bare collinear nonsinglet parton distribution.
Because of the light-cone separation $z$ involved in its definition,
$M^\bare_{+}$ is not directly computable on a Euclidean lattice. We can define a
different quantity that is amenable to lattice simulations  by choosing a purely
spatial separation, $z=\left(0,0,0,z_3\right)$, together with
$\gamma^{\mu}=\gamma^{0}$ and $P=\left(E,0,0,P_3\right)$. Then taking the time
component of Eq.~\eqref{eq:Ioffedec} we get
\begin{align}
	\label{eq:pseudoIoffe}
	M^\bare_0\left(z, P\right) 
	= 2 E\,\mathcal{M}^{(0)}\lp \nu,-z_3^2 \rp \, .
\end{align}
The correlators defined in Eqs.~\eqref{eq:barePDF} and~\eqref{eq:pseudoIoffe}
are known in the literature as (bare) \textit{Ioffe-time} distribution (ITD) and
pseudodistribution (pseudo-ITD)
respectively~\cite{Radyushkin:2017cyf,Braun:1994jq}. 
For $z_3^2  \neq 0$, 
in addition to usual ultraviolet (UV) divergences (leading to  coupling renormalization), 
they   have specific  link-related UV divergences, 
which are regularized by a finite lattice spacing $a$. 
Thus, $\mathcal{M}^{(0)}\lp \nu,-z_3^2 \rp$ is in fact $\mathcal{M}^{(0)}\lp \nu,-z_3^2; a^2 \rp$.

The $a \to 0$ UV divergences are multiplicatively renormalizable~\cite{Ji:2017oey,Ishikawa:2017faj}. 
The relevant renormalization factor $Z(z_3^2, a^2)$ does not depend on $\nu$ and, 
 for small $z_3^2$,   is known at one loop. 
Its explicit form is inessential if one introduces the  
so-called reduced 
Ioffe-time pseudo-distributions
 first defined in Ref.~\cite{Radyushkin:2017cyf}
as 
\begin{align}
	\label{eq:reducedIoffe}
	\mathfrak{M}\left(\nu, z_3^2\right) = \frac{\mathcal{M}\lp \nu,-z_3^2; a^2 \rp}{\mathcal{M}\lp 0,-z_3^2; a^2 \rp}. 
\end{align}
The $Z$-factors of the numerator and denominator are the same and cancel in the ratio leaving the reduced distribution on the left-hand side without any residual dependence on 
the lattice spacing.

Working in the small-$z_3^2$ limit, the pseudo-ITD can be matched at one-loop level to the corresponding ITD
through a finite perturbative kernel,
expressing the pseudo-ITD in terms of the collinear PDFs through a factorization formula based on the operator product expansion (OPE).  
The computation of the relevant QCD diagrams has been performed in a number of independent papers.
The original QCD computation is reported, for example, 
in Refs.~\cite{Radyushkin:2017lvu, Radyushkin:2016hsy, Izubuchi:2018srq, Ji:2017rah}.
A simple discussion of the basic features of  the derivation of the factorization formula in non-gauge theories can be found in Ref.~\cite{DelDebbio:2020cbz}.

The  QCD  result reads
\begin{align}
	\label{eq:factIoffe}
	\mathfrak{M}\left(\nu, z_3^2\right) = \int_{-1}^{1} dx\,C\left(x\nu,\mu^2 z_3^2\right)f\left(x,\mu^2\right) + \mathcal{O}\left(z_3^2\Lambda^2\right)\,, 
\end{align}
with
\begin{align}
	\label{eq:WilsonCoeffIoffe}
	C\left(\xi,\mu^2 z_3^2\right) 
	= e^{i\xi} \nonumber -\frac{\alpha_s}{2\pi}& C_F \int_0^1 dw \, \biggl[\frac{1+w^2}{1-w} \log\left(z_3^2\mu^2\frac{e^{2\gamma_E + 1}}{4}\right) \nonumber \\
	&+4\frac{\log\left(1-w\right)}{1-w} -2\left(1-w\right)\biggr]_+ e^{i \xi w} + \mathcal{O}\left(\alpha_s^2\right).
\end{align}
Eqs.~\eqref{eq:factIoffe}, ~\eqref{eq:WilsonCoeffIoffe} allow to relate collinear PDFs to quantities
which are computable in lattice QCD simulations, through a factorized expression similar to those
relating collinear PDFs to physical cross sections. In the spirit of the ``good lattice cross sections''
proposed in Refs.~\cite{Ma:2017pxb, Ma:2014jla}, this formula can be used in a fitting framework, to extract
PDFs from lattice data, performing the same kind of analysis which is usually done when considering experimental data.
This approach was first studied in Ref.~\cite{Karpie2019}, and subsequently in Ref.~\cite{Izubuchi:2019lyk,Cichy2019,Gao:2020ito}. In Ref.~\cite{Cichy2019}, it has been implemented
within the {\tt NNPDF } fitting environment. Considering data for quasi-PDFs matrix elements 
produced in Refs.~\cite{Alexandrou:2018pbm, Alexandrou:2019lfo} and
starting from the momentum space factorization formula connecting quasi-PDFs to collinear PDFs,
upon numerical implementation of the Fourier transform an expression analogous to Eq.~\eqref{eq:factIoffe}
was obtained, relating parton distributions directly to position space quasi-PDFs matrix elements. A similar analysis was very recently performed by the JAM collaboration in Ref.~\cite{Bringewatt:2020ixn} for the spin-averaged and spin-dependent PDFs employing quasi-PDF lattice data.  

In the present work we perform an analogous exercise considering this time the reduced pseudo-ITD approach,
implementing within the {\tt NNPDF} environment the lattice data presented in Ref.~\cite{Joo:2019jct,Joo:2020spy},
and using the position space factorization formula of Eqs.~\eqref{eq:factIoffe}, \eqref{eq:WilsonCoeffIoffe}.
This, besides being a complementary exercise to the one performed in Ref.~\cite{Cichy2019}, has also some
practical advantages. First, when working in the pseudo-ITD approach, the factorization is realized in the limit of small-$z^2$.
Unlike in the quasi-PDFs approach, where the factorization is realized for high values of $P$,
here we are allowed to keep in the analysis data coming from a wide range of momentum values, 
without having to remove those with lower $P$.
This advantage is particularly important, because in lattice QCD, the low momentum data are significantly more precise for a fixed computational cost.
Second, we can directly use the position space factorization formula of Eq.~\eqref{eq:factIoffe}, relying on the analytical
expression for the perturbative coefficient of Eq.~\eqref{eq:WilsonCoeffIoffe} and without having
to perform the numerical Fourier transform described in the appendix A of Ref.~\cite{Cichy2019}.

In this article we extend the general strategy that has been developed within the {\tt NNPDF} framework and which allows us to systematically extract parton distributions from the available lattice data. In the implementation of this idea once the lattice data have reached some level of maturity in terms of precision and systematic effects, one could combine data from all pertinent lattice formalisms such as quasi-distributions ~\cite{Lin:2014zya,Chen:2016utp,Alexandrou:2015rja,Alexandrou:2016jqi, Monahan:2016bvm, Zhang:2017bzy, Alexandrou:2017huk, Green:2017xeu,Alexandrou:2018pbm,Chen:2018xof,Alexandrou:2018eet,Lin:2018qky,Fan:2018dxu,Liu:2018hxv,Alexandrou:2019lfo,Izubuchi:2019lyk,Green:2020xco,Chai:2020nxw,Lin:2020ssv,Alexandrou:2020zbe}
and pseudo-distributions ~\cite{Orginos:2017kos,Karpie:2018zaz,Joo:2019jct,Joo:2019bzr,Joo:2020spy,Bhat:2020ktg,Fan:2020cpa,Gao:2020ito}.
One can also include results from the so-called ``Good Lattice Cross-Sections'' (LCS) approach,
which is described in~\cite{Ma:2017pxb} and represents a general framework, where one computes matrix elements that can be factorized into PDFs at short distances. 
Papers~\cite{Bali:2017gfr,Bali:2018spj,Sufian:2019bol,Bali:2019ecy, Sufian:2020vzb} describe implementations of the latter formalism.
Clearly a global analysis only makes sense after having scrutinised each set of data individually, and having understood the systematics that affect them.
The structure of the paper is as  follows.  In Sec.~\ref{sec:data} we define the lattice observable considered in the fit,
 describe the corresponding data and  briefly recall the main features of the NLO terms entering the factorization formulas.
In Sec.~\ref{sec:fit} we present the first set of results:
we consider the  fits where only the statistical uncertainties of the lattice data are taken into account.
Analyzing data from different lattice ensembles we 
show that, in general, without accounting for systematic effects  it is not possible to obtain a good fit.
In Sec.~\ref{sec:sys_fits} we discuss and quantify some of the systematic uncertainties affecting
the lattice data.  We  include them into the analysis and  study their impact on the final PDFs and on the fit quality.
Sec.~\ref{sec:conclusions} summarizes our conclusions. 

%% file: sect2.tex
\section{Lattice data and observables}
\label{sec:data}

In this section we describe the lattice observables we will consider in the present work, together with the corresponding data.
By lattice observable, we mean a quantity which can be computed on the lattice on one hand, and related to some
collinear PDFs through some kind of factorization theorem on the other. 
We will consider two different observables corresponding to 
the real and imaginary part of the reduced pseudo-ITD defined in Eq.~\eqref{eq:reducedIoffe}.
 
\begin{figure}[h!]
    \center
    \includegraphics[width=12cm,height=5cm,keepaspectratio]{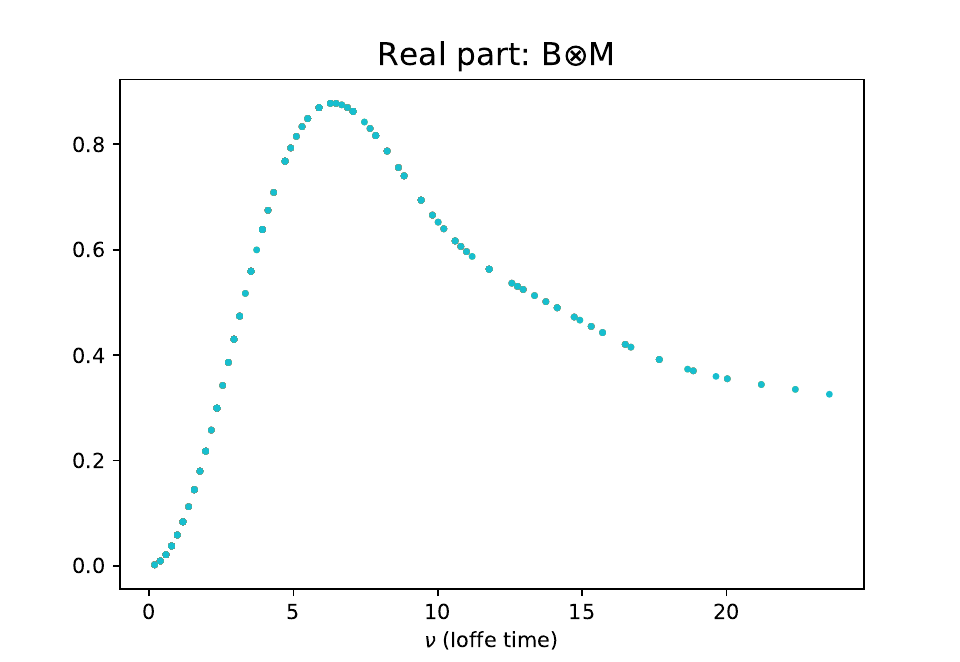}
    \includegraphics[width=12cm,height=5cm,keepaspectratio]{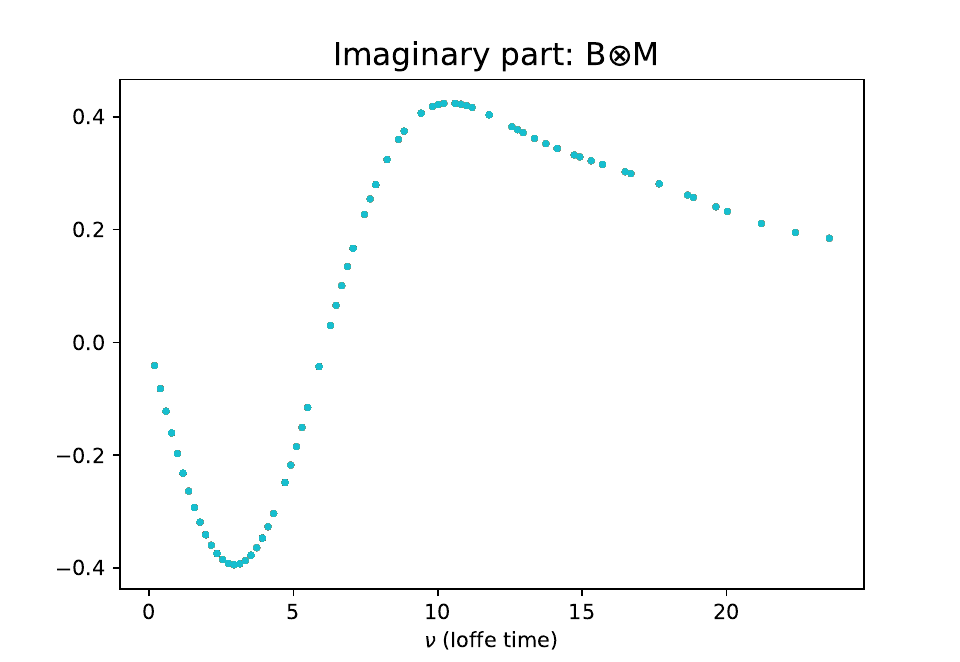}
    \includegraphics[width=12cm,height=5cm,keepaspectratio]{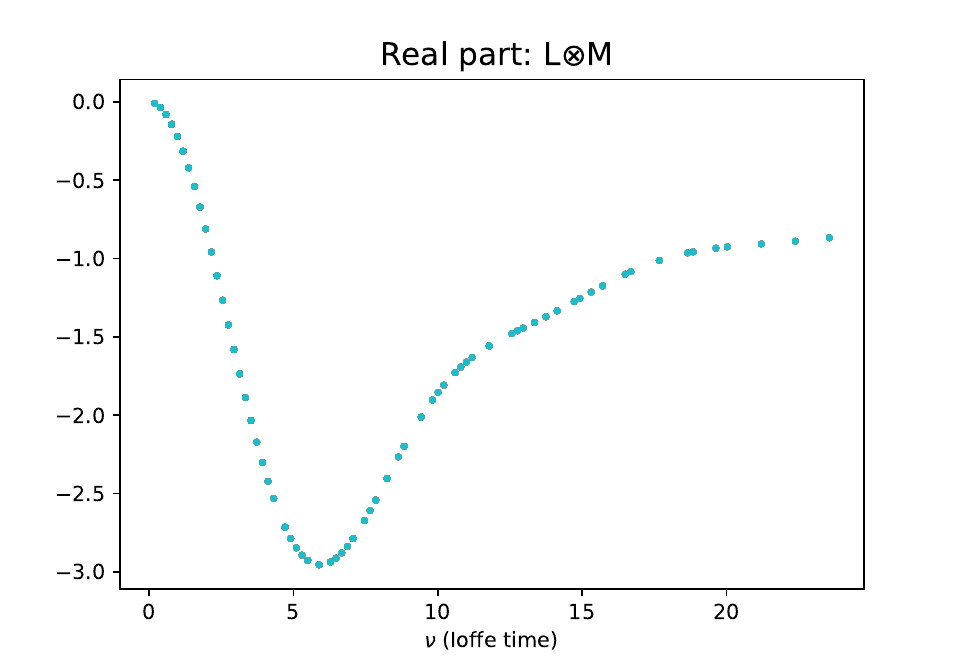}
    \includegraphics[width=12cm,height=5cm,keepaspectratio]{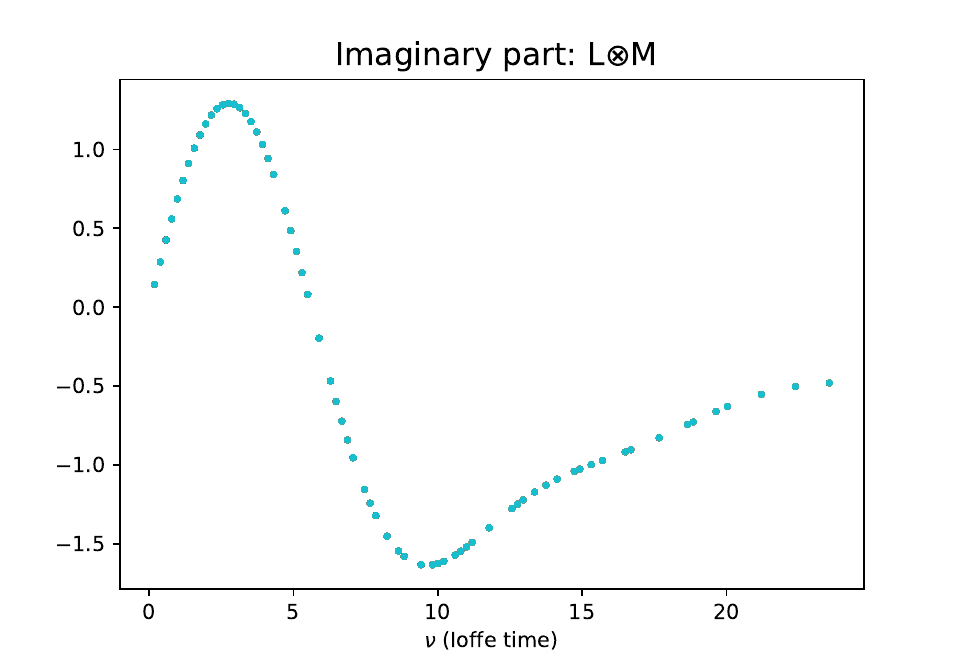}
    \caption{Upper plot: The NLO evolution term for the real (left) and imaginary part (right). 
    Lower plot: The NLO scheme change term for the real (left) and imaginary part (right).}
    \label{fig::BL}
\end{figure}

Considering the case of the unpolarized isovector parton distribution and recalling the definition of the two nonsinglet PDFs
$V_3$ and $T_3$ 
\begin{align}
	&V_3\left(x\right) = u\left(x\right) - \bar{u}\left(x\right) - \left[d\left(x\right) -\bar{d}\left(x\right)\right],  \\
	&T_3\left(x\right) = u\left(x\right) - \bar{u}\left(x\right) + \left[d\left(x\right) -\bar{d}\left(x\right)\right],
\end{align}
taking the real and complex parts of Eq.~\eqref{eq:factIoffe} and using Eq.~\eqref{eq:WilsonCoeffIoffe}, 
we can define the two lattice observables
\begin{align}
	\label{eq:Reppdf}
    &\text{Re}\left[\mathfrak{M}\right]\left(\nu, -z_3^2\right) 
    = \int_{0}^{1} dx\,C^{\text{Re}}\left(x\nu,\mu^2 z_3^2\right)V_3\left(x,\mu^2\right)\, ,\\
	\label{eq:Imppdf}
    &\text{Im}\left[\mathfrak{M}\right]\left(\nu, -z_3^2\right) 
    = \int_{0}^{1} dx\,C^{\text{Im}}\left(x\nu,\mu^2 z_3^2\right)T_3\left(x,\mu^2\right)\, ,
\end{align}
with
\begin{align}
	\label{eq:ReC}
	&C^{\text{Re}}\left(\xi,\mu^2 z_3^2\right) 
    = \cos\left(\xi\right) -\frac{\alpha_s}{2\pi} C_F \int_0^1 dw \, \biggl[B\left(w\right) 
    \log\left(z_3^2\mu^2\frac{e^{2\gamma_E + 1}}{4}\right) + L\left(w\right)\biggr] \cos\left(\xi w\right), \\
    \label{eq:ImC}
	&C^{\text{Im}}\left(\xi,\mu^2 z_3^2\right) 
    = \sin\left(\xi\right) -\frac{\alpha_s}{2\pi} C_F \int_0^1 dw \, \biggl[B\left(w\right)
    \log\left(z_3^2\mu^2\frac{e^{2\gamma_E + 1}}{4}\right)  + L\left(w\right)\biggr] \sin\left(\xi w\right)\, ,
\end{align}
where the kernels $B\left(w\right)$ and $L\left(w\right)$, 
according to Eq. (\ref{eq:WilsonCoeffIoffe}), are given by 
\begin{align}
    \label{eq:B}
    &B\left(w\right) = \left[\frac{1+w^2}{1-w}\right]_+ \, ,\\
    \label{eq:L}
    &L\left(w\right) = \left[4\frac{\log\left(1-w\right)}{1-w} -2\left(1-w\right)\right]_+ \, .
\end{align}
It is worth recalling some important features of the NLO coefficients given in Eqs.~\eqref{eq:ReC},~\eqref{eq:ImC}.
The contributions proportional to the two kernels $B\left(w\right)$ and $L\left(w\right)$ of Eqs.~\eqref{eq:B},~\eqref{eq:L} 
can be seen as an evolution and a scheme change term respectively \cite{Joo:2019jct,Radyushkin:2018cvn}: 
while the former is responsible for the evolution from the PDF scale 
$\hat{z}^{-2} = \mu^2\frac{e^{2\gamma_E + 1}}{4} $ to the pseudo-ITD scale $z^2$, the latter takes into
account the finite terms characterizing the specific choice of the renormalization scheme. 
They are plotted in Fig.~\ref{fig::BL} for both the real and imaginary part, using the PDFs set 
NNPDF31\_nlo\_as\_0118 as input.
The evolution term $B\left(w\right)$ also connects pseudo-ITD points having different values of $z^2$: 
considering for example the real part, from Eqs.~\eqref{eq:Reppdf},~\eqref{eq:ReC} it follows
\begin{align}
    \label{eq::evolRe}
        \text{Re}\left[\mathcal{M}\right]\left(\nu, z_0^2\right) &= 
        \text{Re}\left[\mathcal{M}\right]\left(\nu, z^2\right) \nonumber\\ 
        &- C_F \frac{\alpha_s}{2\pi}\log\frac{z_0^2}{z^2} 
		\int_{0}^{1} dx\,\left[\int_0^1 dw \,B\left(w\right)  \cos\left(x\nu w\right)\right] 
		V_3\left(x,\mu^2\right)\,,
\end{align}
which relates the real part of the pseudo-ITD point at the scale $z^2$ with the one
having the same Ioffe time at the scale $z_0^2$~\cite{Orginos:2017kos,Karpie:2017bzm}. 

In the present work, we will consider the data for reduced pseudo-ITD from Refs.~\cite{Joo:2019jct, Joo:2020spy}:
the datasets presented in Ref.~\cite{Joo:2019jct} have been produced starting from three different lattice 
ensembles, denoted as \textit{fine}, \textit{big} and \textit{coarse} and which differ for the volume and lattice spacing 
used in the simulations. They have been produced using values of the pion mass ranging from 358 MeV (\textit{fine})
to 415 MeV (\textit{coarse} and \textit{big}). 
In the present work we will focus on the datapoints produced from the \textit{fine} ensemble, 
while those from the \textit{coarse} and \textit{big} ones will be used to estimate systematic effects due to continuum limit 
and finite lattice volume.
We will also consider pseudo-ITD points presented in Ref.~\cite{Joo:2020spy}, produced using pion mass equal to 172 MeV.
Following the original convention of Ref.~\cite{Joo:2020spy} we will denote the corresponding lattice ensemble as \textit{170}.
Points from the ensemble \textit{280}, presented in the same paper and produced using similar lattice spacing and pion mass 278 MeV,
will be used to estimate the pion mass effects in the analyses for the ensembles fine and 170.
These five ensembles of $2+1$ flavor lattice QCD were generated by the JLab/W\&M collaboration using clover Wilson fermions 
and a tree level tadpole-improved Symanzik gauge action. One iteration of stout smearing with the weight $ \rho = 0.125$ 
for the staples is used in the fermion action. 
A direct consequence of the stout smearing is that the value of the tadpole corrected tree-level clover coefficient $c_{\rm SW}$ 
used is very close to the non-perturbative value determined, a posteriori, using the Schr\"odinger functional method.
The detailed features of these ensembles are reported in Tab.~\ref{tab:data},
together with the number of reduced pseudo-ITD datapoints $n_{\text{dat}}$ computed from each of them.
\begin{table}[t]
	\renewcommand*{\arraystretch}{1.60}
	\scriptsize
	\centering
	\input{tables/data.tex}
	\vspace{0.3cm}
	\caption{Lattice data details}
	\label{tab:data}
\end{table}

Given a set of lattice data for the real and imaginary part of the reduced pseudo-ITD, 
the distributions $T_3$ and $V_3$ can be extracted from them through a standard minimum-$\chi^2$ fit,
following the approach described in Refs.~\cite{Cichy2019,DelDebbio:2020cbz}: 
the unknown $x$-dependence of the PDFs is parameterized at the chosen scale $\mu^2$, 
using a suitable parametric form, whose best parameters are determined minimizing the $\chi^2$ built using
Eqs.~\eqref{eq:Reppdf},~\eqref{eq:Imppdf} and the corresponding lattice results. In this work we will perform
this exercise using the {\tt NNPDF} fitting framework, running the same machinery commonly used to extract PDFs from experimental data,
already applied to lattice results in Ref.~\cite{Cichy2019}.
In the following we briefly recall its main relevant features, referring to Ref.~\cite{Cichy2019} for more details.

The $x$-dependence of the distributions $f_q\left(x\right)$ ($V_3$ and $T_3$ in our case) 
is parameterized through a neural network $\text{NN}_q$ multiplied by a preprocessing polynomial factor, as
\begin{align}
    \label{eq:NN}
    f_q\left(x\right) = x^{\alpha_q}\left(1-x\right)^{\beta_q}\text{NN}_q\left(x\right)\, ,
\end{align}
$\alpha_q, \beta_q$ being additional free parameters to be determined during the fit, alongside
the weights and biases defining the neural network. Denoting the free parameters of the model as $\theta$, 
the best fit is determined minimizing the $\chi^2$ function, defined as
\begin{align}
	\label{eq::chi2}                   
    \chi^2\left(\theta\right) = \frac{1}{N_{\text{dat}}}\sum_{i,j}
    \left(\mathcal{O}\lp z_i \rp-\mathcal{O}^{\text{th}}\lp z_i, \theta \rp \right)
    \left[\text{Cov}^{-1}\right]_{\text{ij}}
    \left(\mathcal{O}\lp z_j \rp-\mathcal{O}^{\text{th}}\lp z_j, \theta \rp \right), 
\end{align}
where $\mathcal{O}\lp z_i \rp $ denotes the measured lattice observable and
$\mathcal{O}^{\text{th}}\lp z_i, \theta \rp$ is the corresponding theoretical
prediction, expressed using the matching coefficients of Eqs.~\eqref{eq:ReC}, \eqref{eq:ImC} and the parameterized parton
distribution of Eq.~\eqref{eq:NN}. The implementation of the convolution entering Eqs.~\eqref{eq:Reppdf},~\eqref{eq:Imppdf}
is performed by means of FastKernel tables, introduced and validated in Refs. \cite{Ball:2010de,Bertone:2016lga}
in the context of global QCD fits, and currently used within the {\tt NNPDF} code to obtain all the required theoretical 
predictions in a global fit.
The covariance matrix entering Eq.~\eqref{eq::chi2} describes the distribution of the data, and 
takes into account the statistical and systematic uncertainties and their correlations.
Considering $N_c$ independent sources of correlated systematics, its explicit expression is given by 
\begin{align}
    \label{eq:covariance}
    \left[\text{Cov}\right]_{\text{ij}} = \delta_{ij}\left(\sigma^{\text{stat}}_{i,s}\right)^2 
    + \sum_{l=1}^{N_c}\sigma^{\text{sys}}_{i,l}\sigma^{\text{sys}}_{j,l}\,,
\end{align}
where $\sigma^{\text{stat}}_{i}$ and $\sigma^{\text{sys}}_{i,l}$  represent the statistical and the l-th correlated 
systematic uncertainty of the i-th point.
 
The covariance matrix defined in Eq.~\eqref{eq:covariance} enters both the definition of the $\chi^2$  
and the generation of Monte Carlo replicas \cite{Ball:2014uwa},
being therefore important for both the central value of the fit and the final PDFs error. A solid knowledge 
of the covariance matrix is therefore an essential ingredient to get reliable results.
The minimization of the $\chi^2$ is performed numerically: different algorithms can be implemented,
here the CMA algorithm \cite{DBLP:journals/corr/Hansen16a} is used, employing a cross-validation technique to avoid overfitting. The specific code used in the present work is the one employed for the production of the PDFs set NNPDF31 \cite{Ball:2017nwa}, together with the {\tt ReportEngine} software \cite{zahari_kassabov_2019_2571601}.

The {\tt NNPDF } methodology has been used to produce PDF sets for many years now, and provides a flexible environment
within which it has been possible to fit more than 4000 experimental points, coming from a variety 
of different high energy processes in different kinematic ranges \cite{Ball:2017nwa, Ball:2014uwa}.
Therefore it represents a reliable framework which can be used to study and analyze the available lattice data, to assess
how well these are able to constrain the PDFs and to compare lattice results with those coming from standard PDF sets.
It is important to emphasise once again that in this analysis, once the FastKernel tables have been generated, 
the lattice data are treated exactly on the same footing as any other data, viz. the exact same methodology 
and code are used for fitting experimental and lattice data.

%% file: tables/data.tex
\begin{tabularx}{\textwidth}{XXXXXccccc}
    \toprule
     Lattice ensemble        &     a(fm)  &   $M_{\pi}$ (MeV)   &  $L^3 \times T$ & $n_{\rm dat}$ & Reference   \\
    \midrule
     fine                         &  0.094(1)  &  358(3)  &  $32^3\times 64$ &       48  &    \\
     big                      &  0.127(2)  &  415(23)  &  $32^3\times 96$ &       48   &  \cite{Joo:2019jct,Joo:2020spy}  \\
     coarse                      &  0.127(2)  &  415(23)  &  $24^3\times 64$ &       36   &   \\
     \midrule
     280                        &  0.094(1)  &  278(3)   &  $32^3\times 64$ &       64    &  \cite{Joo:2020spy}  \\
     170                         & 0.091(1)  &  172(6)   &  $64^3\times 128$&       80   &   \\
    \bottomrule
    \end{tabularx}
    

%% file: sect3.tex
\section{Fits over lattice data: statistical uncertainties only}
\label{sec:fit}

In this section, we will present results for fits performed over the lattice data
computed from the ensembles \textit{fine} and \textit{170}, denoted as \textit{fine-stat} and \textit{170-stat}
respectively. 
Such fits have been produced considering statistical uncertainties only. We will show how, in general,
without having the complete information regarding the lattice systematic uncertainties 
it is not always possible to obtain a good fit.
In the next section, taking as example the case of the \textit{fine} ensemble, we will discuss and estimate some of the possible
systematic effects, studying their impact on the fit quality and on the resulting PDFs. 

\begin{figure}[h!]
    \centering
    \includegraphics[width=12cm,height=5cm,keepaspectratio]{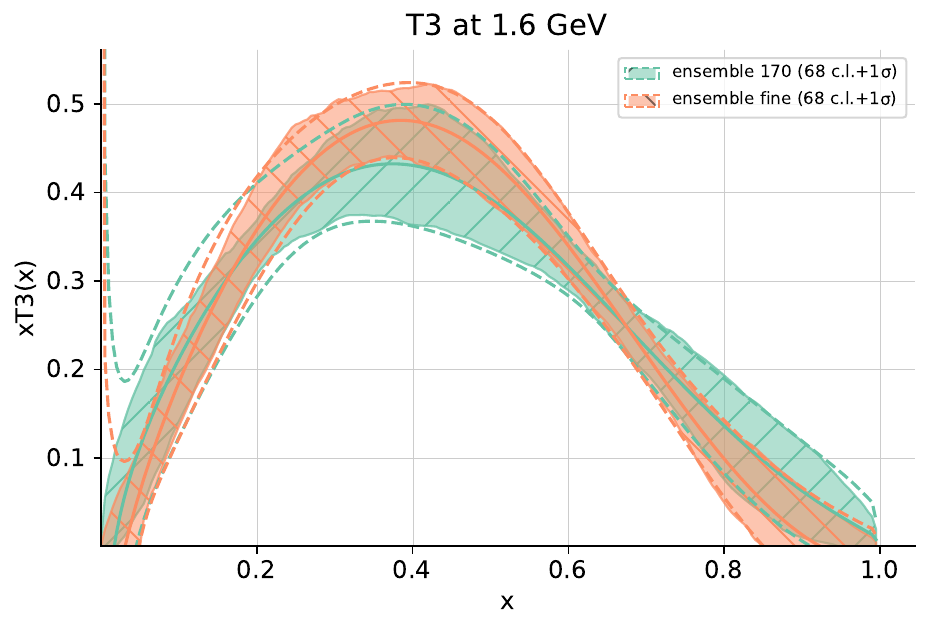}  
    \includegraphics[width=12cm,height=5cm,keepaspectratio]{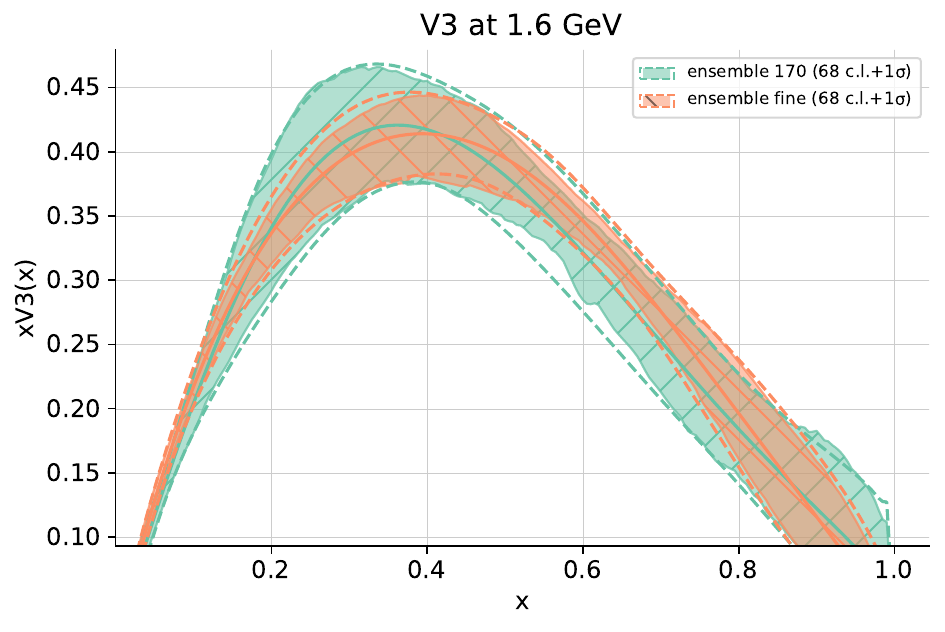} 
    \includegraphics[width=12cm,height=5cm,keepaspectratio]{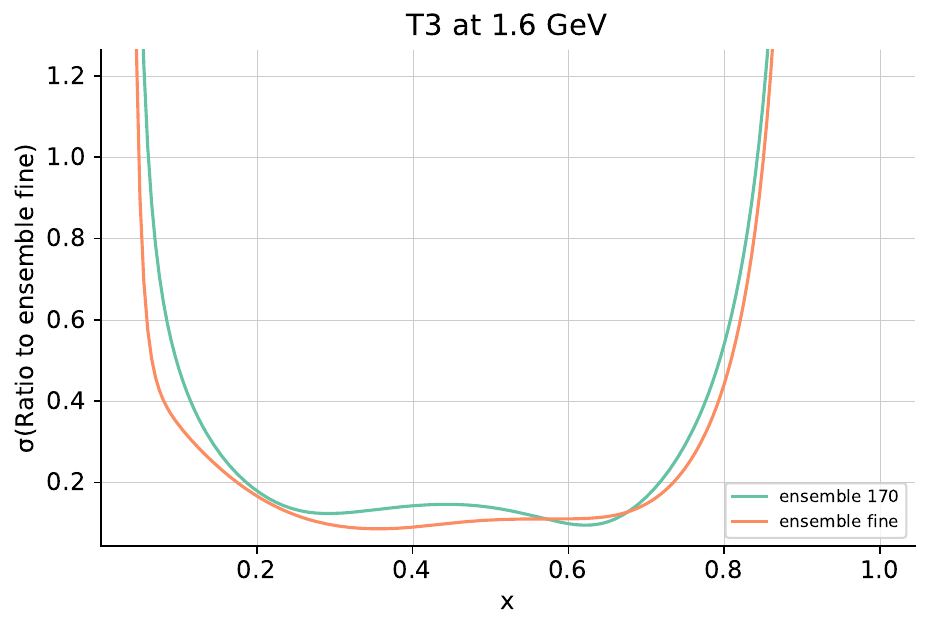}  
    \includegraphics[width=12cm,height=5cm,keepaspectratio]{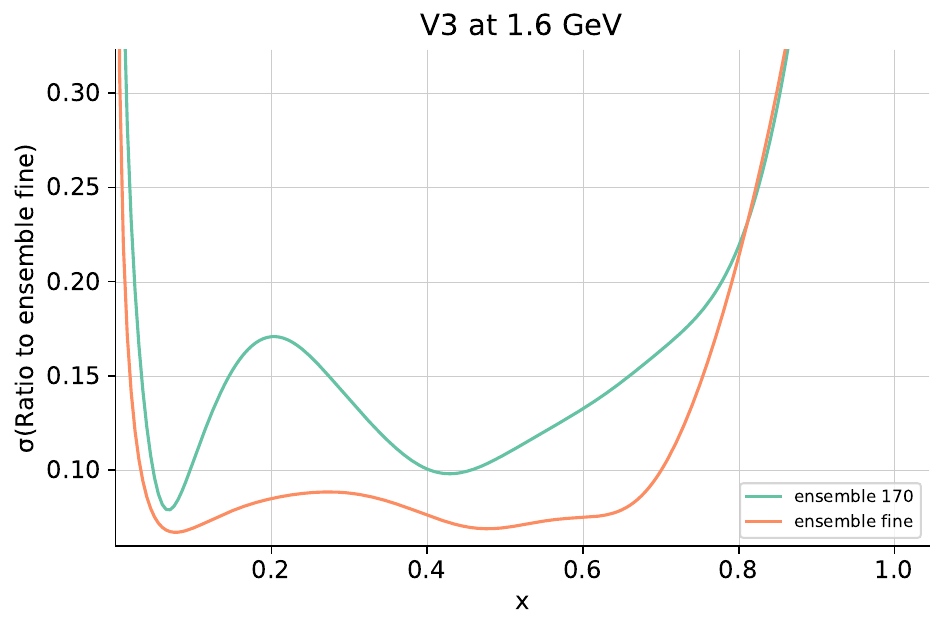}     
\caption{Upper plots: PDFs from datapoints computed from the ensembles fine and 170.
The shaded bands represent the PDFs error computed as the 68 c.l. of the fit replicas, while the dashed line is obtained by computing the standard deviation point by point in $x$.
Lower plots: corresponding PDFs errors, computed as standard deviation over fit replicas and 
displayed in function of $x$.}

\label{fig:pdfs_fine_170}
\end{figure}

Parton distributions resulting from fits \textit{fine-stat} and \textit{170-stat}, together with the corresponding error bands, are plotted in the upper and lower
plots of Fig.~\ref{fig:pdfs_fine_170}, and the $\chi^2$ values are reported in Tab.~\ref{tab:chi2_stat_fine_170}:  
despite the PDFs extracted from the two datasets are compatible within one $\sigma$, the error band
of the fit \textit{fine-stat} appears to be slightly  smaller than the other, with an average $\chi^2$ value per datapoint 
equal to $8.36$, pointing out a possible underestimation of the error and a bad fit quality.
This could be caused by inconsistencies between different datapoints, due to unknown systematic uncertainties affecting them.
On the other hand, the fit \textit{170-stat} shows better $\chi^2$ values, with an average value per datapoint equal to $1.38$.

Focusing on the more problematic case of the fine ensemble results,
in order to assess which points are more likely to be affected by large systematic errors, 
we will study the contribution to the $\chi^2$ coming from each datapoint
\begin{align}
	\label{eq:contribution}
		\delta_i = \frac{\left(D_i - T_i\right)^2}{\sigma_i^{2}}\, , 
\end{align}
$D_i$ and $T_i$ being the $i$-th lattice point and the corresponding prediction from the fit respectively,
and find out which points $D_i$ are more than $4 \sigma $ (or $3 \sigma $) off from the fitted distribution $T_i$.
These are the points that, most likely, do not belong to the fitted distribution 
and which therefore might be affected by larger systematic effects.

\begin{table}[t]
	\renewcommand*{\arraystretch}{1.60}
	\scriptsize
	\centering
	\input{tables/chi_stat.tex}
	\vspace{0.3cm}
	\caption{Details of fits with statistical uncertainties only. 
	From left to right we report the lattice ensemble, the fit name, 
	the observables included in the analysis, the number of datapoints and finally
	the partial and total $\chi^2$.}
	\label{tab:chi2_stat_fine_170}
\end{table}

\begin{figure}[h!]
    \center
    \includegraphics[width=12cm,height=5cm,keepaspectratio]{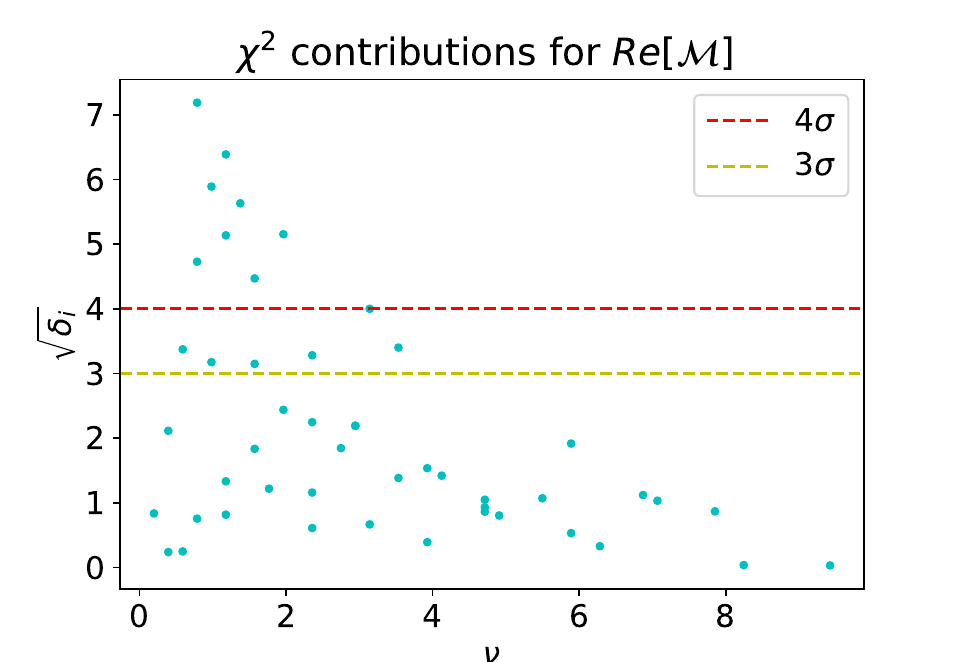}  
    \includegraphics[width=12cm,height=5cm,keepaspectratio]{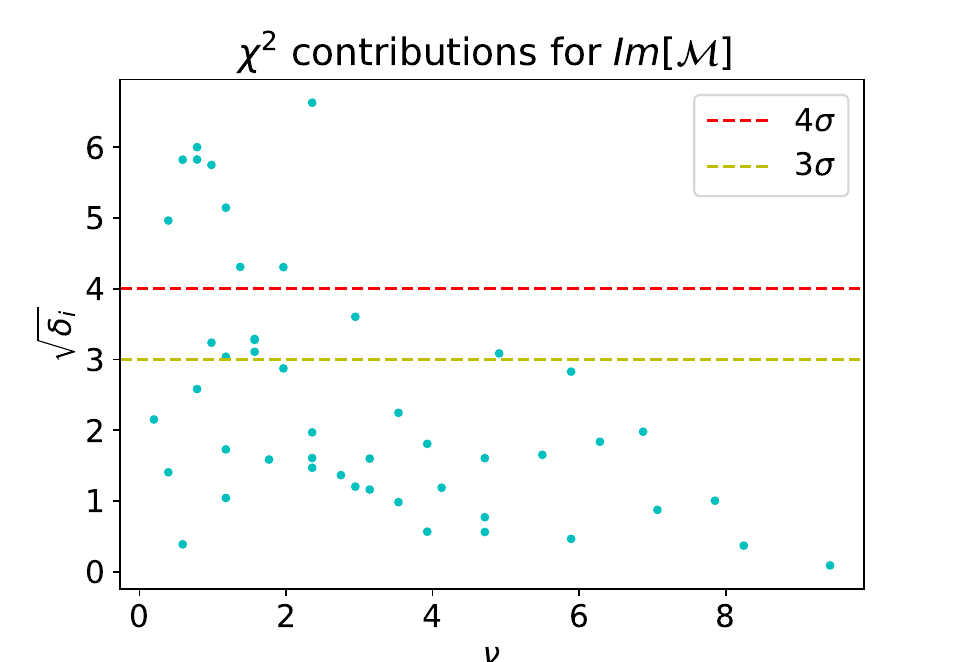}
    \includegraphics[width=11.5cm,height=4.5cm,keepaspectratio]{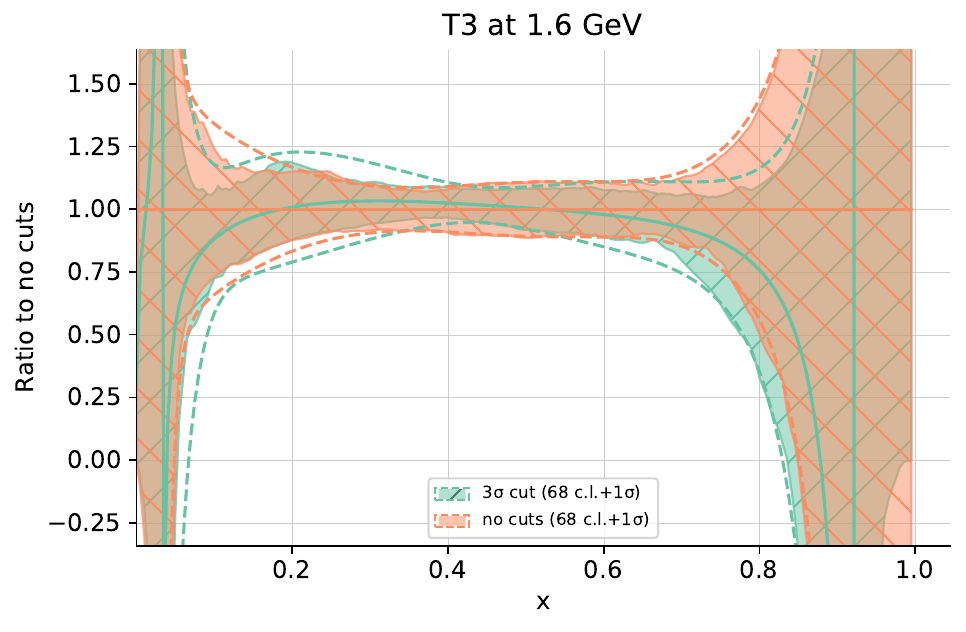}  
    \includegraphics[width=11.5cm,height=4.5cm,keepaspectratio]{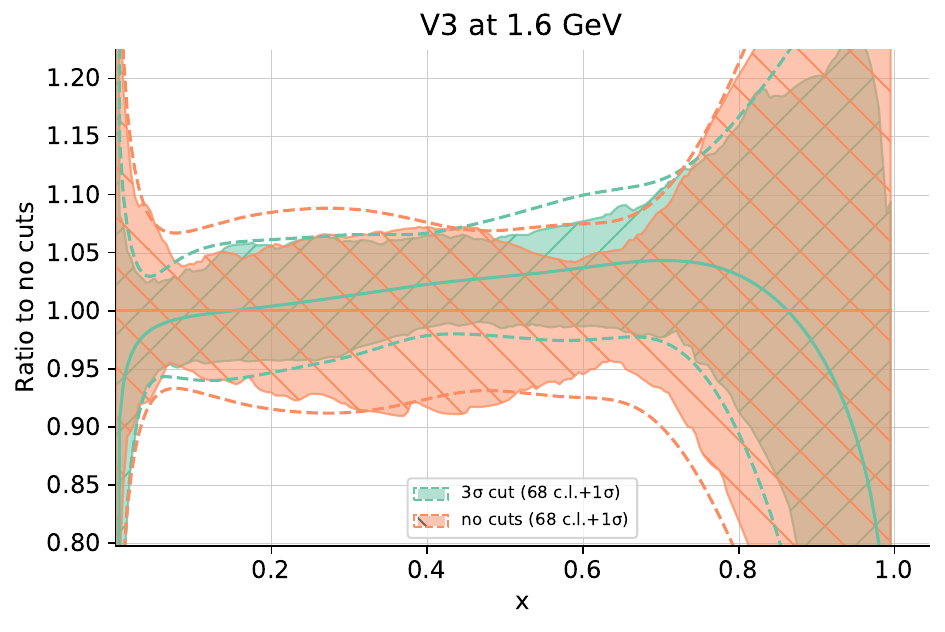}
\caption{Upper plots: $\sqrt{\delta_i}$ contributions for each datapoint of the fine ensemble. The red and
yellow lines highlight the $4\sigma$ and $3\sigma$ cut respectively. Lower plots: 
PDFs from fits \textit{fine-stat} (orange) and \textit{fine-stat-3$\sigma$} (green), normalized to the former.}
\label{fig:chi_contributions}
\end{figure}

The contributions $\sqrt{\delta_i}$ are plotted in the upper plot of Fig.~\ref{fig:chi_contributions} as a function of the
Ioffe-time $\nu$, with the red and yellow lines highlighting the $4\sigma$ and $3\sigma$ cut respectively:
it is clear that a bunch of points having small Ioffe-time values
are those giving the highest contribution to the total $\chi^2$, being more than $3\sigma$ or $4\sigma$ off.
We can implement $4\sigma$ and $3\sigma$ cuts, removing the problematic points from the dataset and
producing new fits, denoted as \textit{fine-stat-$3\sigma$} and \textit{fine-stat-$4\sigma$}: 
the new fits show more reasonable $\chi^2$ values, reported in Tab.~\ref{tab:chi2_stat_fine_170},
showing how, upon removing the outliers, the remaining points, coming from a wide range of momentum $p$
and Euclidean separation $z_3$, are fitted reasonably well.
The PDFs resulting from the $3\sigma$ cut are plotted in the lower plot of Fig.~\ref{fig:chi_contributions},
normalized to the fit without any cuts: it is clear how, despite spoiling the total $\chi^2$,
the problematic points do not seem to have a big impact on the final PDFs.

We conclude that, depending on the specific lattice ensemble we consider, quite a high number of small Ioffe-time points 
do not belong to the fitted distribution.
In order to get reasonable $\chi^2$ values, such points have to be removed from the fit.
This highlights possible tensions between datapoints and may point out the presence 
of systematic effects.
In order to avoid any underestimation of the PDFs error and to introduce back in the analysis all the available points, 
systematic uncertainties need to be quantified and implemented in the fit.

%% file: tables/chi_stat.tex
\begin{tabularx}{\textwidth}{XXXXXXXcccccc}
\toprule
 Ensemble & fit         & Obs            & $n_{\rm dat}$ & $\chi^2$  & $\chi^2_{tot}$  \\
\midrule
fine    & fine-stat       & $\text{Re}\left[\mathfrak{M}\right]$      &  48           & 7.94      & 8.36  \\
        &               & $\text{Im}\left[\mathfrak{M}\right]$      &  48           & 8.77      &       \\
        & fine-stat-$4\sigma$ & $\text{Re}\left[\mathfrak{M}\right]$      &  39           & 2.68      & 3.28  \\
        &               & $\text{Im}\left[\mathfrak{M}\right]$      &  39           & 3.89      &       \\ 
        & fine-stat-$3\sigma$  & $\text{Re}\left[\mathfrak{M}\right]$      &  34           & 1.45      & 1.86  \\
        &               & $\text{Im}\left[\mathfrak{M}\right]$      &  32           & 2.27      &       \\
\midrule
170    & 170-stat       & $\text{Re}\left[\mathfrak{M}\right]$       &  80           & 0.68      & 1.38  \\
       &               & $\text{Im}\left[\mathfrak{M}\right]$       &  80           & 2.07      &       \\

\bottomrule
\end{tabularx}

%% file: sect4.tex
\section{Systematic effects}
\label{sec:sys_fits}

\subsection{Discussion}
\label{sec:discussion_sys}

The high $\chi^2$ values of the fits presented in the previous section might point out
the presence of some tensions between datapoints.
In the following, focusing on the case of the fine ensemble results, we will show that this is indeed the case, 
and we will investigate possible sources of systematic uncertainties and their numerical values.

\begin{figure}[h!]
    \center
    \includegraphics[width=12cm,height=5cm,keepaspectratio]{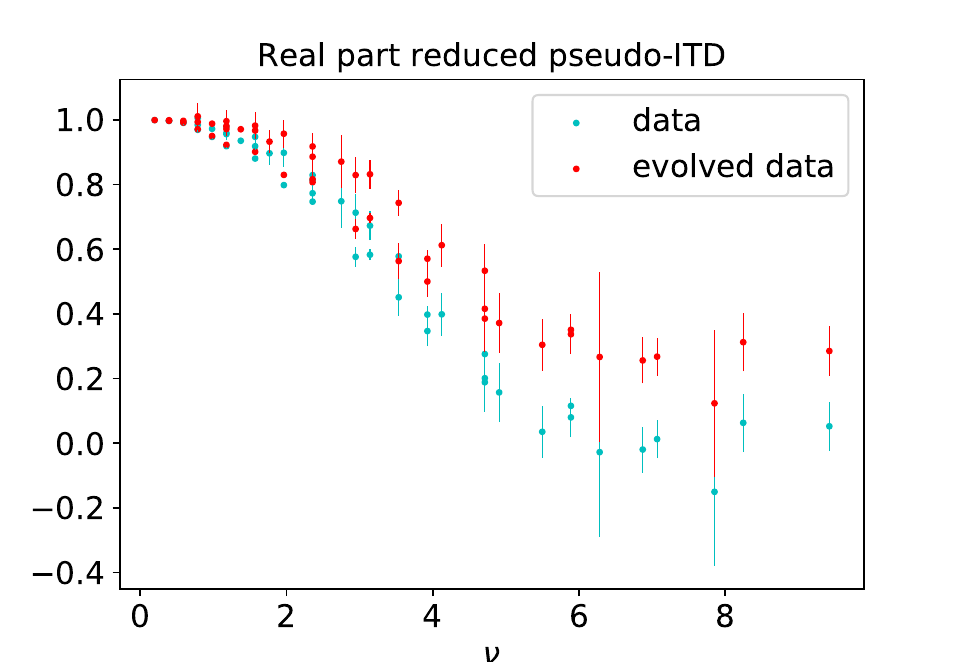}
    \includegraphics[width=12cm,height=5cm,keepaspectratio]{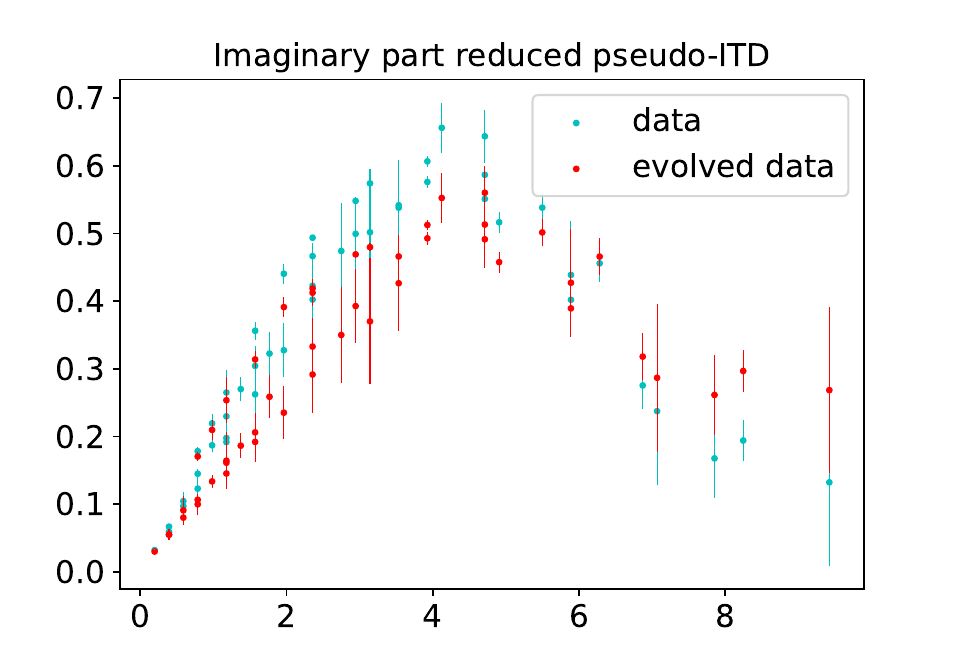}
    \caption{Data for the real part of the pseudo-ITD at their original scale $z^2$ and evolved at the common scale $z_0^2$.}
    \label{fig::evol}
\end{figure}

The matrix element defining the pseudo-ITD is a function of the Ioffe-time $\nu$ and of the scale $z^2$.
Points having the same Ioffe-time but different Euclidean separation can be related through Eq.~\eqref{eq::evolRe}, 
which can be used to evolve each pseudo-ITD point up to a chosen reference scale $z_0 ^2 = \left(0.7\,a\right)^2$.
Looking at Fig.~\ref{fig::BL} it is clear that, given this choice for $z_0$, the sign of the NLO correction 
of Eq.~\eqref{eq::evolRe} will be positive for every datapoint, so that the evolution increases the real part of the pseudo-ITD.
Considering the imaginary part, the sign of the NLO evolution term is initially negative, and it turns positive at bigger values of $\nu$.
Such effects can be seen in Fig.~\ref{fig::evol}, where the pseudo-ITD points computed from the fine ensemble are plotted
before (blue) and after evolution (red). 
After evolution, points having the same Ioffe time should have the same value.
In practice, they should be compatible within errors.
Looking at the red points of Fig.~\ref{fig::evol}, where each point is plotted with the corresponding statistical uncertainty, 
it is clear how, expecially in the small Ioffe time region, this is not always the case:
after evolution, some points having the same Ioffe time are not compatible between each other.
Such discrepancies might be explained by the presence of systematic effects we are not accounting for.

A proper investigation of the systematic effects affecting the computation of the equal time correlators underlying
the definition of pseudo-PDFs is a difficult and expensive task which would require to run different lattice simulations 
varying a set of parameters, like for example the lattice spacing, the lattice volume, the pion mass. 
Alongside systematic effects due to the lattice simulation, other sources of errors are those connected 
to the theoretical framework of the pseudo-PDFs approach, like the presence of higher twist effects and perturbative matching truncation effects. 
A detailed discussion of many of these uncertainties, together with a series of possible scenarios for their numerical values, 
can be found for example in Ref.~\cite{Cichy2019}. 

\begin{figure}[h!]
    \centering
    \includegraphics[width=12cm,height=5cm,keepaspectratio]{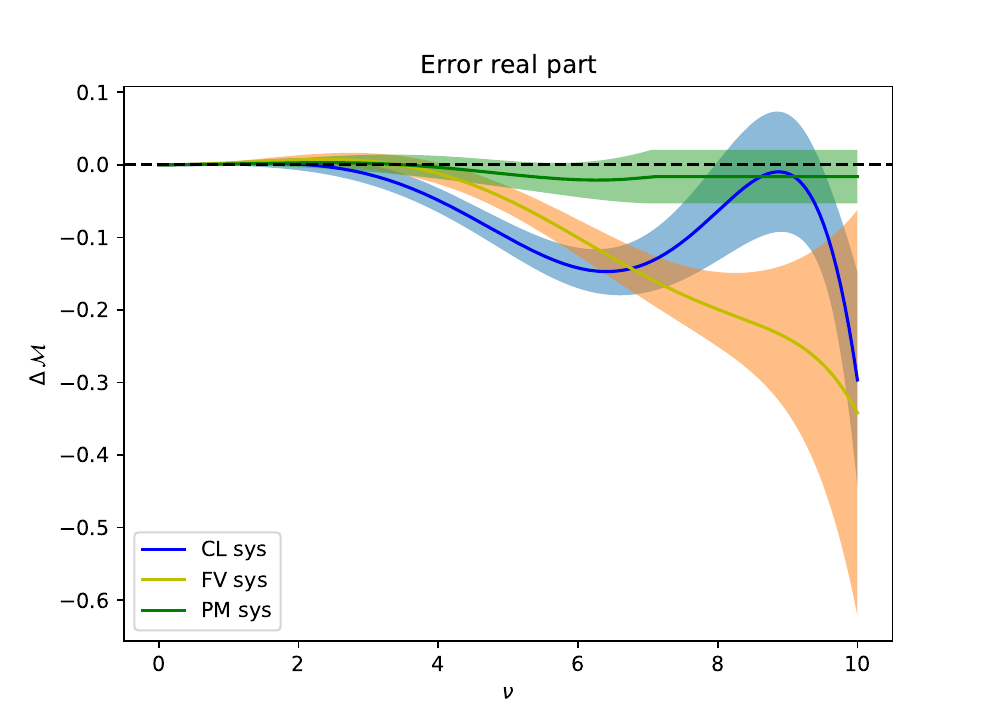}  
    \includegraphics[width=12cm,height=5cm,keepaspectratio]{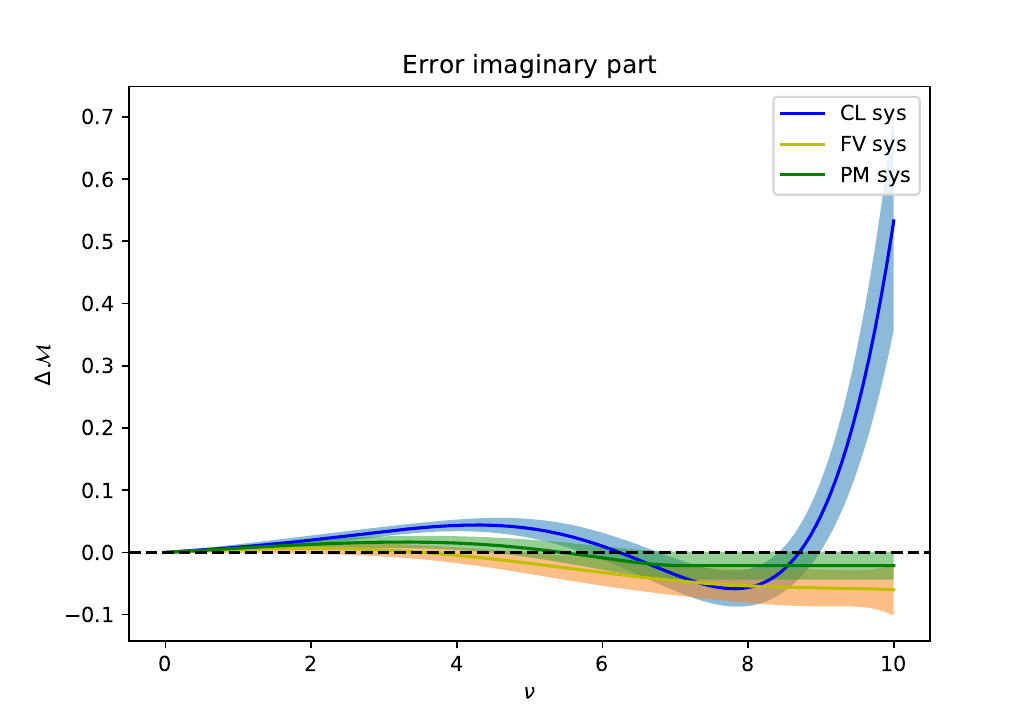}
    \includegraphics[width=12cm,height=5cm,keepaspectratio]{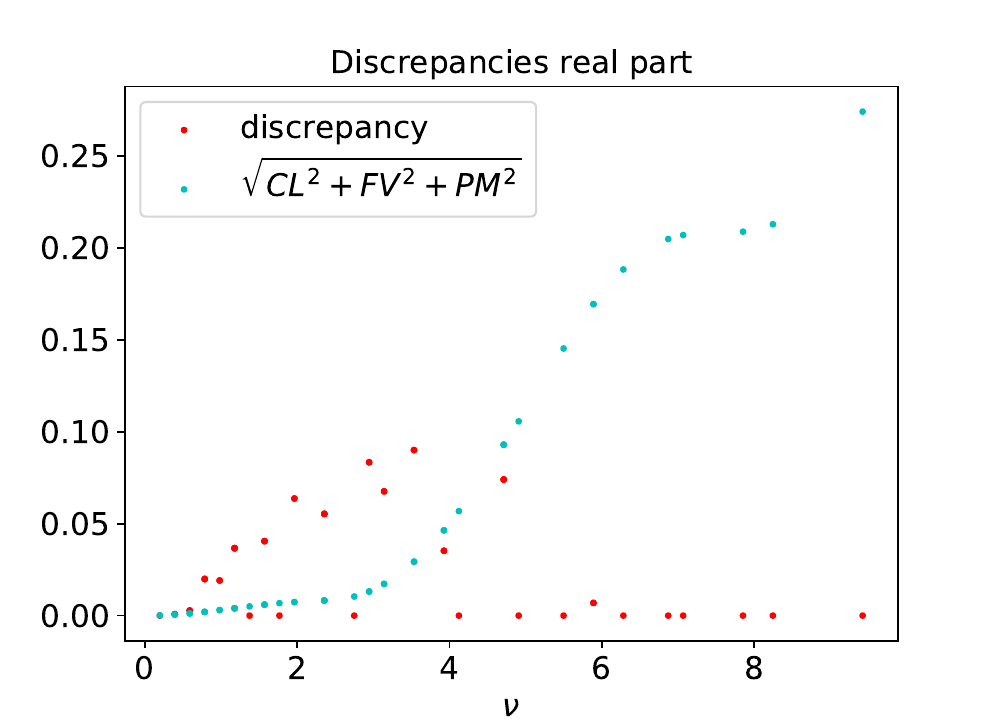}  
    \includegraphics[width=12cm,height=5cm,keepaspectratio]{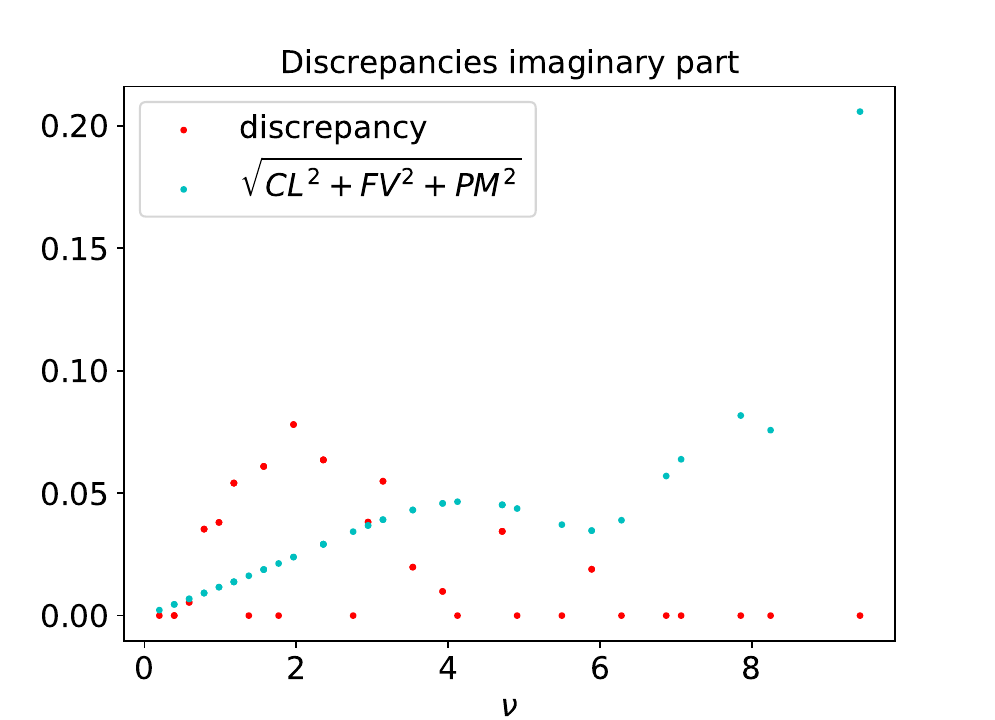}
\caption{Upper plots: finite volume (FV), continuum limit (CL) and pion mass (PM) systematics provided 
as functions of the ioffe-time $\nu$ for the real (left) and imaginary (right) part 
of the matrix element. Lower plots: discrepancies between data having the same ioffe-time (red)
together with the total FV, CL and PM systematic effects (blue).}
\label{fig:sys}
\end{figure}

As mentioned in Sec.~\ref{sec:data},
in Ref.~\cite{Joo:2019jct} additional pseudo-ITD points were computed starting
from other two lattice ensembles, with pion mass similar to that of the fine one, but having different volume and lattice spacing, 
denoted as \textit{big} and \textit{coarse}, whose features are reported in Tab.~\ref{tab:data}.
Systematic uncertainties due to the continuum limit (CL) and finite volume (FV) can be directly estimated using 
these additional results as detailed in Ref.~\cite{Joo:2019jct}: the real and imaginary components 
of the pseudo-ITD are fitted to a polynomial as a function of the Ioffe-time $\nu$; 
the difference between coarse and fine ensemble results is taken
as an estimate for lattice spacing effects as a function of $\nu$, while the analogous difference 
considering the coarse and big ensembles gives an estimate for uncertainties due to finite lattice volume.
Systematic effects due to the pion mass (PM) can be estimated in a similar way: 
as mentioned in Sec.~\ref{sec:data}, in Ref.~\cite{Joo:2020spy} the data of the ensembles fine and 170 have been supplemented with
additional pseudo-ITD results produced from a third ensemble having pion mass equal to $278$ MeV, denoted as ensemble
\textit{280}.  
The difference between polynomial fits for the ensembles fine and 280 is taken as an estimate for pion mass effects.
These differences will be considered as three independent sources of correlated systematic, affecting each datapoint
entering the analysis. 
They are shown in the upper plots of Fig.~\ref{fig:sys} as functions of the Ioffe-time, denoted as FV (finite volume),
CL (continuum limit) and PM (pion mass).

It is important to understand whether or not these systematic uncertainties are enough to
account for the discrepancies described at the beginning of the section.
In the lower plots of Fig.~\ref{fig:sys} FV, CL and PM systematic effects are plotted for the relevant Ioffe-time values, 
together with the aforementioned discrepancies. 
Consistently with what observed previously, the latter seem to affect mostly low Ioffe-time points, 
which are also those for which the estimated systematics reach their minimum values.
Therefore from Fig.~\ref{fig:sys} it follows that FV, CL and PM systematics cannot be considered responsible 
for the big contributions to the $\chi^2$ noted in the fits of the previous section. 
In other words, they are likely not enough to account for the observed discrepancies 
affecting low Ioffe-time points.
It should be noted that a study of more than 2 ensembles for each systematic error may be necessary for a more definitive conclusion.

Excited states contaminations might represent another possible source of systematic effects. 
Also missing higher orders in perturbation theory and higher twist effects could in principle be treated
as additional systematic uncertainties. 
Unlike the case of the FV, CL and PM systematic uncertainties discussed above, we cannot estimate
the size of such effects using the current lattice results. One could then follow the approach adopted in Ref.~\cite{Cichy2019},
where different scenarios for the size of such systematics have been considered, and try to draw conclusions about their
impact on the PDFs and on the fit quality. 
Here we will follow a different approach, trying to quantify an additional uncertainty which accounts for the 
unknown missing systematic effects, following a Bayesian approach as detailed in the following.

The figure of merit which is minimized during a Gaussian fit is defined as the probability of the data $D$ given
the model parameters $\theta$, namely the likelihood
\begin{align}
    \label{eq:likelihood}
    \mathcal{P}\left(D|\theta\right) = e^{-\frac{1}{2}\left(D-T\left(\theta\right)\right)^T\,
    \Sigma^{-1}\,\left(D-T\left(\theta\right)\right)}.
\end{align}
where $\Sigma$ is the covariance matrix of the data $D$, accounting for the known
statistical and systematic uncertainties, and $T\left(\theta\right)$ is the theoretical prediction,
function of the model parameters.
If we assume the presence of unknown systematic effect $\Delta$ affecting the datapoints $D$, Eq.~\eqref{eq:likelihood}
can be modified as
\begin{align}
    \mathcal{P}\left(D,\Delta|\theta\right) = 
    e^{-\frac{1}{2}\left(D+\Delta-T\left(\theta\right)\right)^T\,
    \Sigma^{-1}\,\left(D+\Delta-T\left(\theta\right)\right)}.
\end{align}
Assuming a Gaussian prior distribution 
$\mathcal{P}\left(\Delta\right) = \exp\left[-\frac{1}{2}\Delta^T\,\hat{\Sigma}^{-1}\,\Delta\right]$
we can marginalize over $\Delta$ getting
\begin{align}
    \label{eq:modifiedlikelihood}
    \int d\Delta\,\mathcal{P}\left(\Delta\right) \mathcal{P}\left(D,\Delta|\theta\right)   
    \propto 
    e^{-\frac{1}{2}\left(D-T\left(\theta\right)\right)\left(\Sigma+\hat{\Sigma}\right)^{-1}\left(D-T\left(\theta\right)\right)},
\end{align}
which defines the relevant likelihood to be minimized.
Eq.~\eqref{eq:modifiedlikelihood} shows how the presence of unknown systematic effects can be accounted for by 
introducing in the likelihood an additional contribution to the covariance matrix, denoted by $\hat{\Sigma}$, 
which defines the prior probability distribution of these systematics. Its specific definition is of course arbitrary,
and depends on the knowledge of the missing uncertainties we have.

This Bayesian approach, despite not providing a general method 
to estimate the missing systematics, allows to include in the analysis the partial information we may have about them.
It should be noted that Eq.~\eqref{eq:modifiedlikelihood} is obtained under the hypothesis that 
the unknown systematic uncertainty is gaussianly distributed. This is an assumption of the model, often done in the literature,
which allows to greatly simplify the problem. However depending on the specific data considered it might not be the most 
realistic one. In the present work we will work using this gaussian assumption, leaving for future studies
the investigation of different choices for the prior probability distribution. 
A gaussian Bayesian approach has already been applied in different physical problems, when the data are affected 
by unknown sources of systematics:
in the case of global QCD analysis, in Refs.~\cite{AbdulKhalek:2019bux,AbdulKhalek:2019ihb},
a suitable covariance matrix $\hat{\Sigma}$ was defined by mean of scale variations, in order to take into account 
the theoretical error due to missing higher orders, while in Ref.~\cite{Bernal:2018cxc} a similar approach was applied 
to cosmological data.

In our case, we only know the discrepancies observed at the beginning of this section,
not described by continuum limit, finite volume and pion mass effects.
We can look at such discrepancies as an indication of the minimal size of the systematic
effects affecting the data and use them to construct a suitable $\hat{\Sigma}$:
for each couple of points having a given Ioffe-time value, we will define the two corresponding diagonal components
of $\hat{\Sigma}$ as half of the distance between evolved points, setting the off diagonal elements to zero.
Each point sharing the same Ioffe time value with at least another one will therefore be affected by an additional, 
uncorrelated systematic such that, after evolution, datapoints having the same Ioffe-time will be compatible between each other. 
Clearly, this global, uncorrelated systematic will be the dominant one for small Ioffe-time points, 
where most of the problematic points are, while for higher value of $\nu$ lattice spacing, finite volume
and mass effects will dominate.

\subsection{Results}
\label{sec:res}
\begin{figure}[h!]
    \center
    \includegraphics[width=12cm,height=5cm,keepaspectratio]{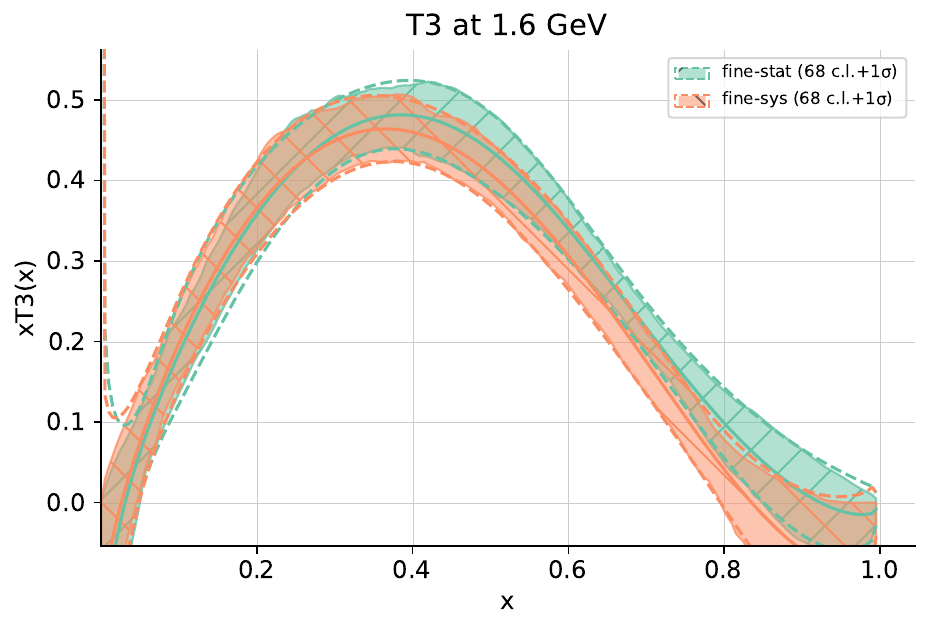}
    \includegraphics[width=12cm,height=5cm,keepaspectratio]{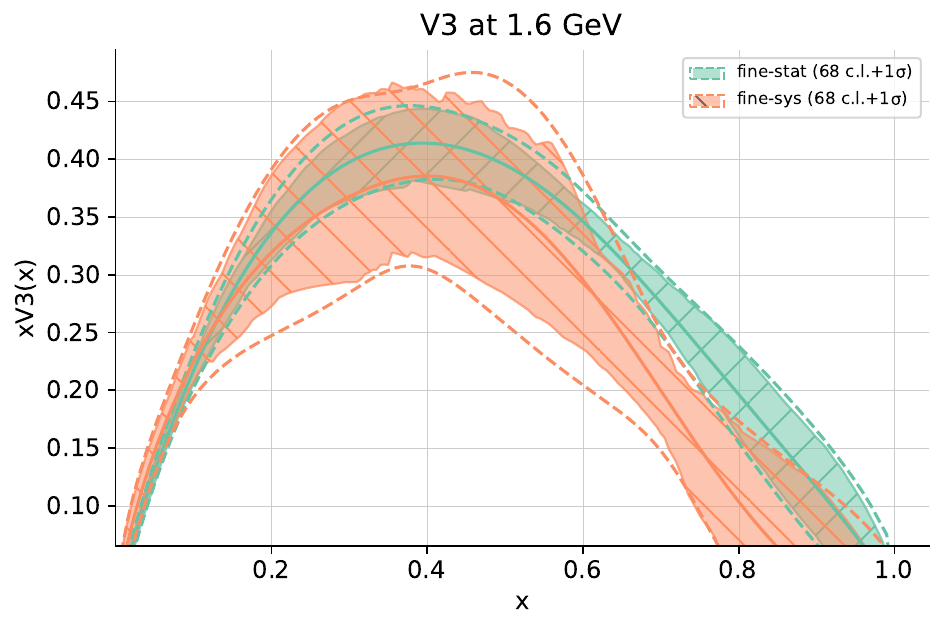}
    \caption{PDFs from the fits \textit{fine-stat} and \textit{fine-sys}.}
    \label{fig::fits_z2}
\end{figure}

To sum up, in Sec.~\ref{sec:discussion_sys}, we have discussed and estimated four different source of systematics: 
the first three, accounting for finite volume, lattice spacing and pion mass effects, can be computed directly from the available 
lattice results as a function of the Ioffe-time $\nu$, and will be implemented in the fit 
as three independent sources of correlated systematics; the fourth one has been estimated using
the size of the discrepancies observed between points having the same Ioffe-time, and will be considered
as an additional uncorrelated uncertainty, in order to take into account the minimal size of all the remaining systematic effects
we have not directly computed. 
As mentioned in Sec.~\ref{sec:data}, such systematics enter the definition of the covariance matrix responsible 
for both replicas generation and the $\chi^2$ definition, and therefore it has a central role in both
the determination of the fit central value and its error band.
The new fit is denoted as \textit{fine-sys} and the resulting PDFs are plotted in Fig.~\ref{fig::fits_z2}, 
together with the results from  the fit \textit{fine-stat} presented Sec.~\ref{sec:fit}:
the distribution $T_3$ is only marginally affected by the introduction of the systematic errors,
showing a mild down shift of its central value in the medium and large $x$ regions; on the other hand 
both the central value and the error band of $V_3$ change, with an overall down shift of the former and 
a sizable increase of the latter. 
The $\chi^2$ values are reported in Tab.~\ref{tab:chi2_sys}: the average value per datapoint is now $1.15$, 
showing a good fit quality.
It should be noted that after the inclusion of systematic uncertainties in the analysis, the effect
on the final result could be different depending on the specific situation we are considering.
In other words, it is not always the case that the inclusion of new systematic effects leads to an increase 
of the final PDFs error. This can be seen for example in the case of the distribution $T_3$ plotted in Fig.~\ref{fig::fits_z2},
from which it is clear how the error of the fit \textit{fine-sys} has not increased with respect to the one of
\textit{fine-stat}. The reason for this can be traced back to the fact that the covariance matrix 
defined in Eq.~\eqref{eq:covariance} enters both the Monte Carlo replicas generation and the $\chi^2$ definition
of Eq.~\eqref{eq::chi2}: while the former mostly controls the final PDFs error, 
the latter is responsible for the relative weights different points have in the analysis.
Points affected by bigger errors will give smaller contributions to the $\chi^2$ and therefore 
will count less in the fit. 
Each replica will be shifted by a certain amount, 
which takes into account both the new replicas distribution and the different weights of the data entering the
$\chi^2$, so that the net effect on the final PDFs is non-trivial, and might consist in a global shift
of the central value of the replicas distribution rather than in an increase of its error band.

\begin{table}[t]
    \renewcommand*{\arraystretch}{1.60}
    \scriptsize
    \centering
    \input{tables/chi_sys.tex}
    \vspace{0.3cm}
    \caption{Details of the fit with systematic uncertainties. 
	From left to right we report the lattice ensemble, the fit name, 
	the observables included in the analysis, the number of datapoints and finally
	the partial and total $\chi^2$.}
    \label{tab:chi2_sys}
\end{table}
    
\begin{figure}[h!]
    \center
    \includegraphics[width=12cm,height=5cm,keepaspectratio]{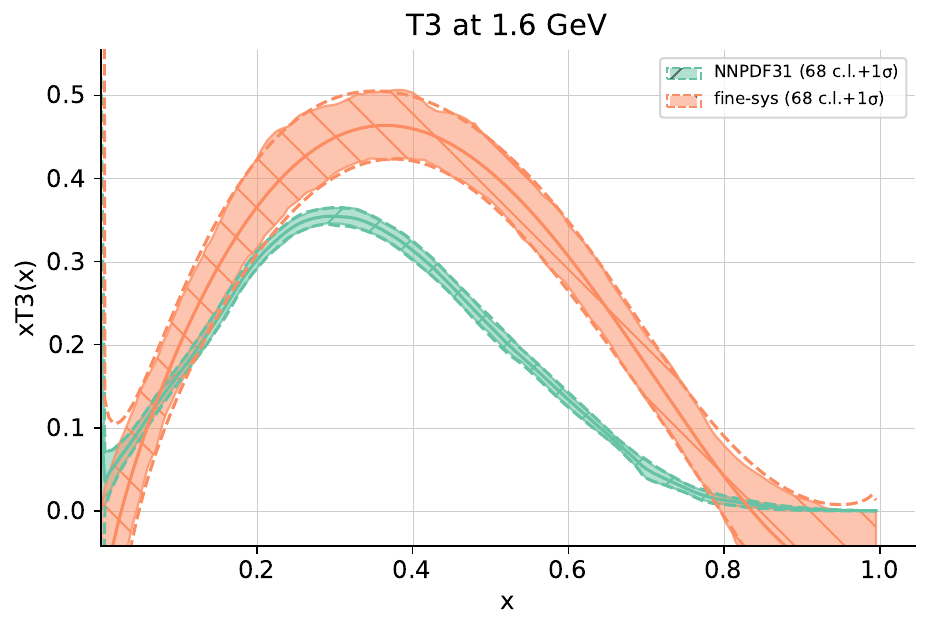}
    \includegraphics[width=12cm,height=5cm,keepaspectratio]{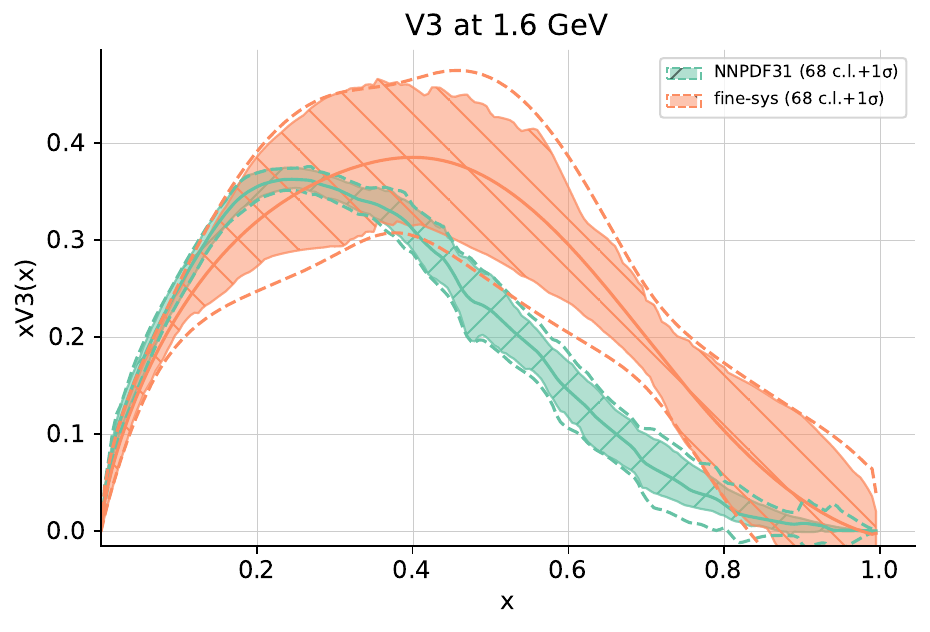}
    \includegraphics[width=12cm,height=5cm,keepaspectratio]{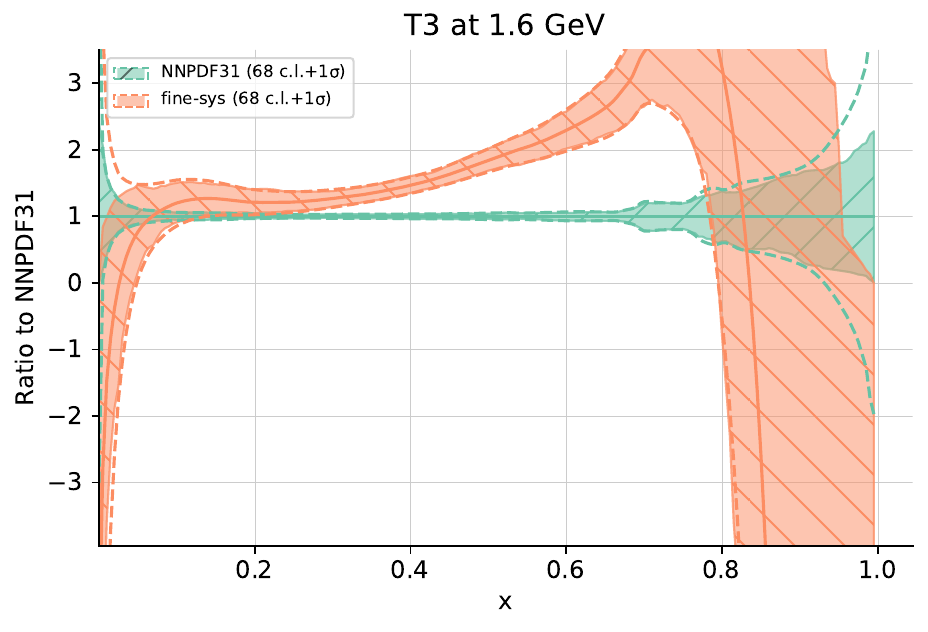}
    \includegraphics[width=12cm,height=5cm,keepaspectratio]{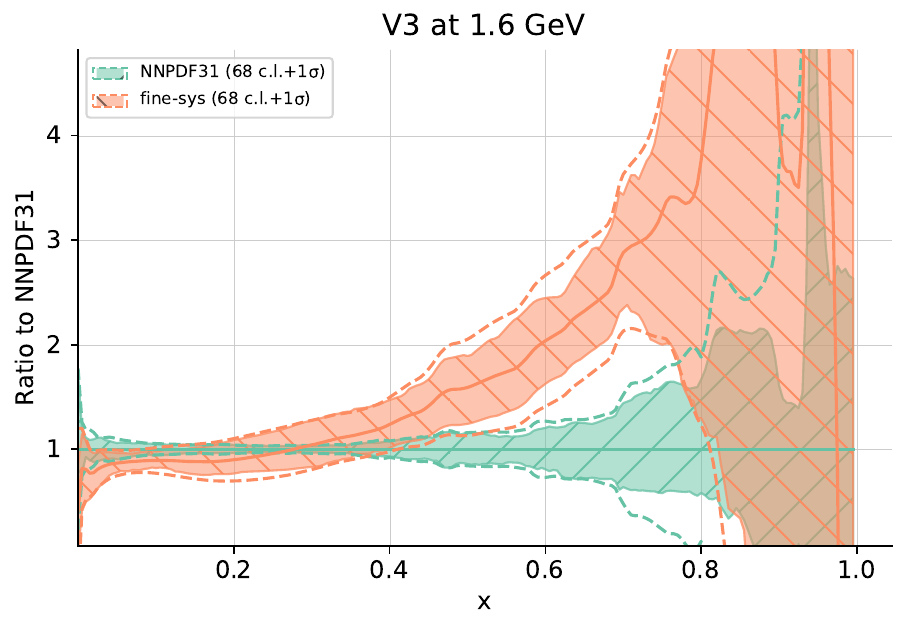}
    \caption{PDFs from the fits \textit{fine-sys} compared with the corresponding distributions from NNPDF31. 
    In the lower plots results are normalized to NNPDF31 PDFs.}
    \label{fig::fits_ratio}
\end{figure}

Despite it is probably too early to draw comparisons between our results and phenomenological distributions, 
it is interesting to see how they look when plotted together: given the fact that nowadays $V_3$ and $T_3$ are
very well constrained by experimental data, the discrepancies we observe between lattice and phenomenological results
might be a good indication of the size of the systematic we are still missing, highlighting specific $x$-region
where the lattice PDFs error might have been underestimated. 
In Fig.~\ref{fig::fits_ratio},
our result \textit{fine-sys} and the corresponding distributions from the NLO PDF sets
NNPDF31 \cite{Ball:2017nwa} are plotted together (orange and green curves respectively), both as
absolute values (upper plots) and normalized to NNPDF31 (lower plots). 
Looking at results from \textit{fine-sys}, in the case of both $V_3$ and $T_3$ the two distributions are compatible 
up to medium ($\sim 0.25$ and $\sim 0.45 $)
and for large values of $x$ ($ >0.8$), showing a probable underestimation of the PDFs error for the intermediate $x$ ranges. 

%% file: tables/chi_sys.tex
\begin{tabularx}{\textwidth}{XXXXXXXcccccc}
\toprule
 Ensemble & fit         & Obs            & $n_{\rm dat}$ & $\chi^2$  & $\chi^2_{tot}$  \\
\midrule
fine    & no cuts       & $\text{Re}\left[\mathfrak{M}\right]$      &  48           & 1.00      & 1.15  \\
        &               & $\text{Im}\left[\mathfrak{M}\right]$      &  48           & 1.30      &       \\

\bottomrule
\end{tabularx}

%% file: sect5.tex
\section{Conclusions}
\label{sec:conclusions}
In the present paper,
we have considered the pseudo-ITD data produced in Refs.~\cite{Joo:2019jct,Joo:2020spy}. Using the
position space factorization theorem relating such data to collinear PDFs, we have extracted two nonsinglet distributions 
within the {\tt NNPDF} framework. 

After extracting PDFs from different data sets and 
considering statistical uncertainties only, 
we have shown that in one of the cases considered, the fit quality appears to be really poor,
pointing out the need for a detailed knowledge of the systematic effects.
Using the results of Ref.~\cite{Joo:2019jct,Joo:2020spy} we have directly estimated those connected to finite volume, lattice spacing
and pion mass effects. As for systematic uncertainties which cannot be directly computed from lattice results (like
for example truncation effects and higher twist corrections), starting from
the observed discrepancies between low Ioffe-time points we have used a Bayesian approach to introduce an additional systematic 
which allows us to mitigate the tensions between the problematic datapoints, using the partial pieces of information which are available to us.

The Bayesian approach however is not completely satisfying, since it relies on a partial knowledge of the
missing uncertainties and requires to make a number of assumptions about them. More work has to be done
to achieve a detailed knowledge of the systematic uncertainties in lattice simulations: 
without a stringent control over them, it is not possible to draw reliable conclusions and
to make comparisons with phenomenological distributions. 

Finally, we stress once more that the analysis performed in this paper is complementary to that 
presented in Ref.~\cite{Cichy2019}, where quasi-PDFs matrix elements where considered instead, 
starting from the momentum space version of the factorization theorem. 
In both cases, results have been produced within the {\tt NNPDF} environment,
running the same machinery used for global QCD analysis over experimental data. 
The next logical step might be a global lattice QCD fit within this same framework, where data for multiple lattice observables 
coming from different simulations are simultaneously included in the analysis.

\section*{Acknowledgements}

We are grateful to E.~R.~Nocera for useful discussions and his interest in this work and to the organisers of ’Parton Distributions and Lattice Calculations’ in Michigan
in September 2019, which has been very useful for the present work.
This work is supported by Jefferson
Science Associates, LLC under U.S. DOE Contract \#DE-AC05-06OR23177.
KO was supported in part by U.S.  DOE grant \mbox{
  \#DE-FG02-04ER41302} and in part by the Center for Nuclear Femtography grants C2-2020-FEMT-006, C2019-FEMT-002-05.
    AR was supported in part by U.S. DOE Grant
\mbox{\#DE-FG02-97ER41028. }  JK was supported
in part by the U.S. Department of Energy under contract
DE-FG02-04ER41302, Department of Energy Office of Science Graduate
Student Research fellowships, through the U.S. Department of Energy,
Office of Science, Office of Workforce Development for Teachers and
Scientists, Office of Science Graduate Student Research (SCGSR)
program and is supported by U.S. Department of Energy grant DE-SC0011941.
TG is supported by The Scottish Funding Council, grant H14027.
The authors gratefully acknowledge the computing time granted by
the John von Neumann Institute for Computing (NIC)
and provided on the supercomputer JURECA at J\"ulich
Supercomputing Centre (JSC)~\cite{JSC}. In addition, this work was made possible using results obtained  at NERSC, a DOE Office of Science User Facility supported by the Office of Science of the U.S. Department of Energy under Contract \mbox{\#DE-AC02-05CH11231}, as well as resources of the Oak Ridge Leadership Computing Facility (ALCC and INCITE) at the Oak Ridge National Laboratory, which is supported by the Office of Science of the U.S. Department of Energy under Contract No. \mbox{\#DE-AC05-00OR22725}.  The software libraries used on these machines were Chroma~\cite{Edwards:2004sx}, QUDA ~\cite{Clark:2009wm,Babich:2010mu}, QDP-JIT~\cite{Winter:2014dka} and QPhiX~\cite{Joo:2013lwm,10.1007/978-3-319-46079-6_30} developed with  
 support from the U.S. Department of Energy, Office of Science, Office of Advanced Scientific Computing Research and Office of Nuclear Physics, Scientific Discovery through Advanced Computing (SciDAC) program, and of the U.S. Department of Energy Exascale Computing Project.

%% file: appendix.tex
\section{Pion Mass dependence for 170 ensemble}
\begin{figure}[h!]
    \centering
    \includegraphics[width=12cm,height=5cm,keepaspectratio]{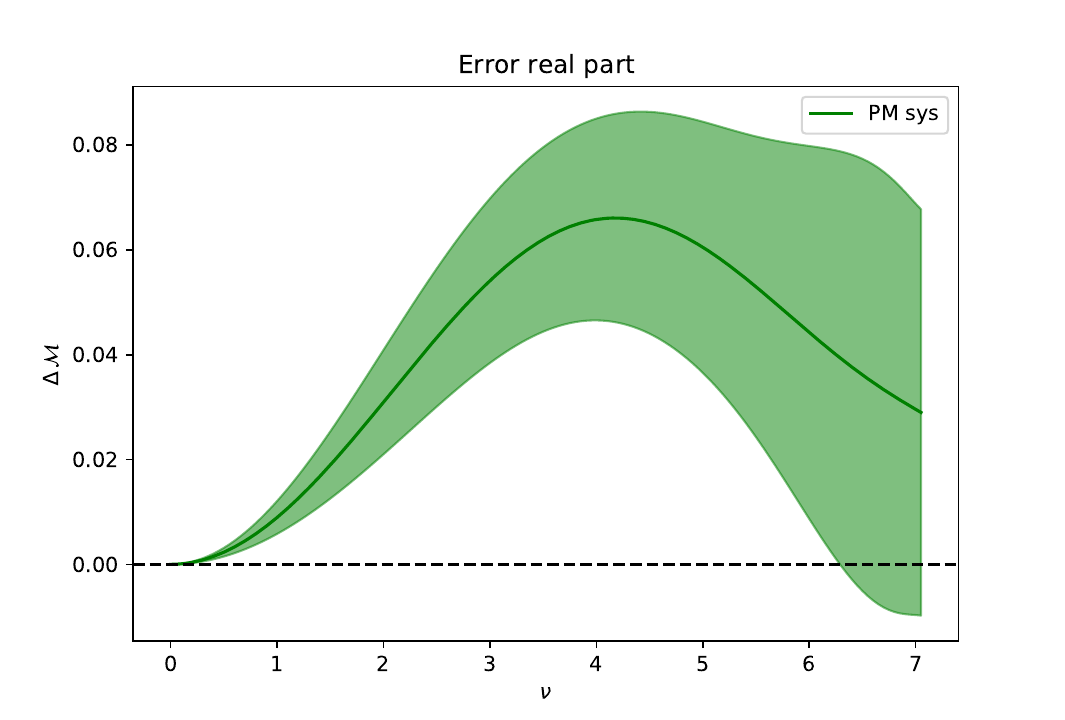}  
    \includegraphics[width=12cm,height=5cm,keepaspectratio]{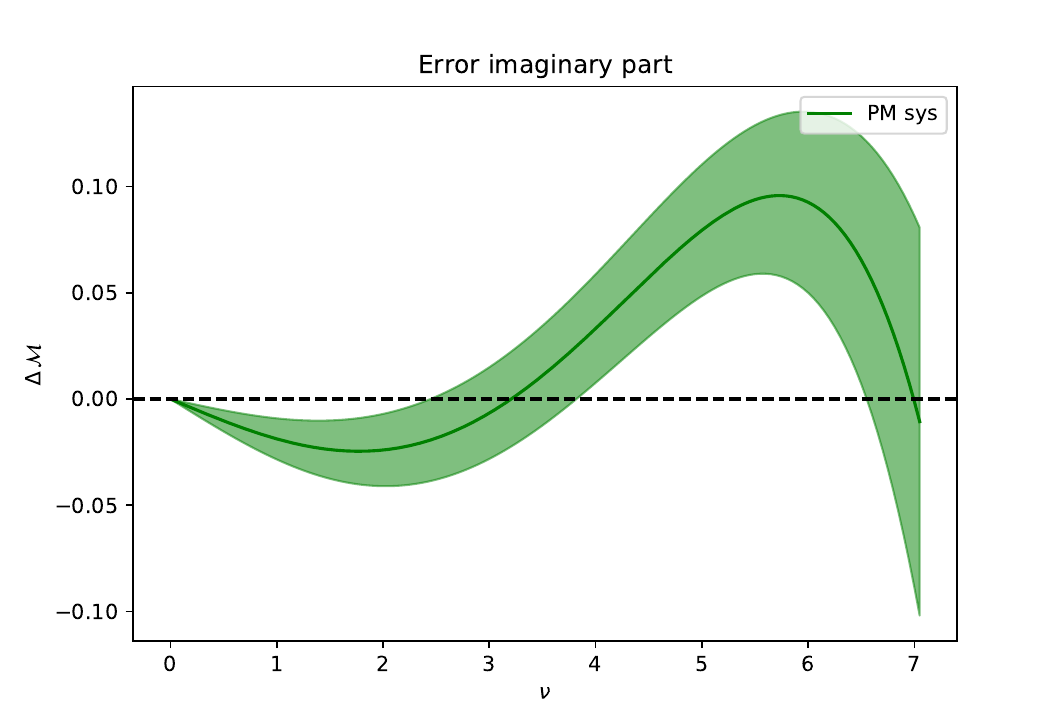}
\caption{Pion mass (PM) systematic provided 
as functions of the ioffe-time $\nu$ for the real (left) and imaginary (right) part 
of the matrix element.}
\label{fig:sys_170}
\end{figure}

Similarly to what done for the fine ensemble in Sec.~\ref{sec:discussion_sys},
the data for the ensemble 280 presented in Ref.~\cite{Joo:2020spy} can also be used to estimate pion mass effects
for results concerning the ensemble 170. The corresponding polynomial curves are plotted in Fig.~\ref{fig:sys_170}
as functions of the Ioffe-time.

\begin{figure}[h!]
    \centering
    \includegraphics[width=12cm,height=5cm,keepaspectratio]{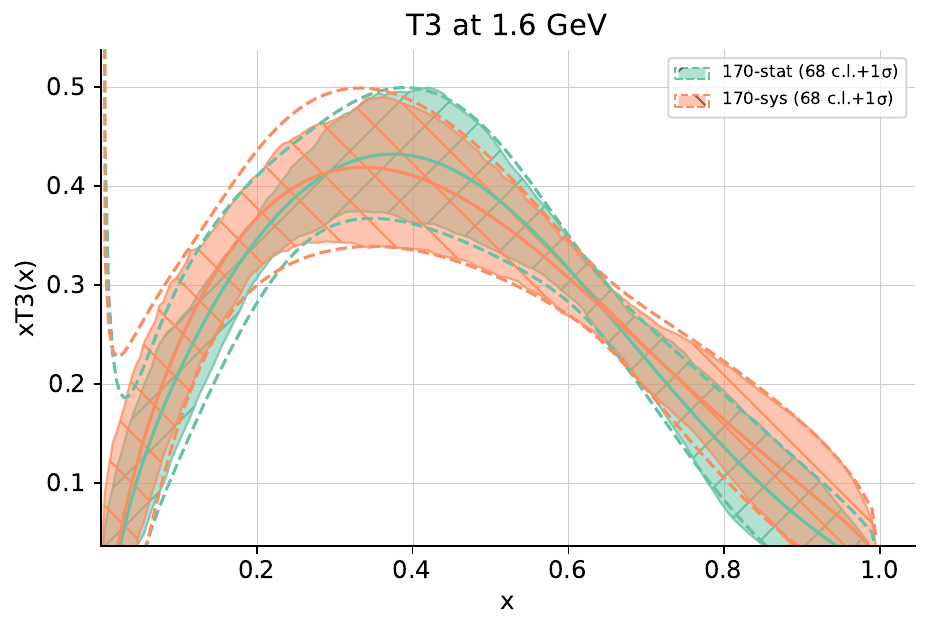}  
    \includegraphics[width=12cm,height=5cm,keepaspectratio]{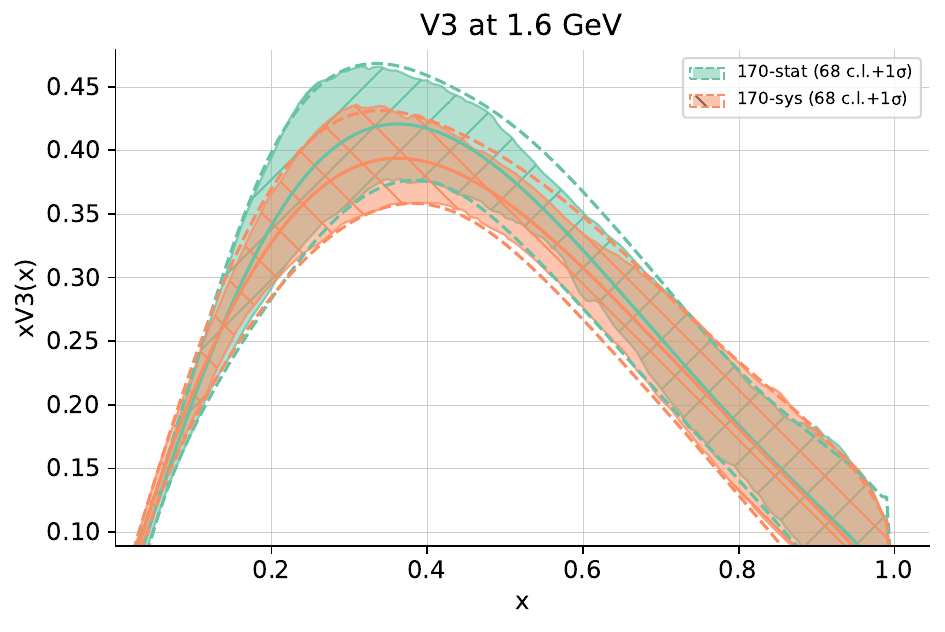}
\caption{PDFs from the fits \textit{170-stat} and \textit{170-sys}.}
\label{fig:res_170_sys}
\end{figure}

As in the case of the analysis for the fine ensemble, the curves in Fig.~\ref{fig:sys_170} are used 
to define a source of correlated systematic. The resulting PDFs, denoted as \textit{170-sys}, are 
plotted in Fig.~\ref{fig:res_170_sys} together with the results for the ensemble 170 presented in Sec.~\ref{sec:fit},
where only statistical uncertainties have been considered.
From Fig.~\ref{fig:res_170_sys} it is clear how introducing pion mass systematic effects in the analysis has very little impact on
the distributions, the major effect being a mild down shift of the central value of $V_3$ in the medium $x$ region.
We conclude that the mild pion mass dependence observed in pseudo-ITD data of Ref.~\cite{Joo:2020spy} has no sizable impact
on the final PDFs.

%% file: appendix2.tex
\section{Comparison with results from quasi-PDFs matrix elements}
It is interesting to compare our best result \textit{fine-sys} with the best result of Ref.~\cite{Cichy2019},
denoted as \textit{nnpdf31\_qpdf\_S2}. 
Both PDFs sets have been obtained using the same NNPDF methodology, the only difference 
being the input data (pesudo-ITD and quasi-PDFs data respectively) and the corresponding errors.
For more details about the specific systematics uncertainties considered in the analysis for \textit{nnpdf31\_qpdf\_S2} we refer 
to the original publication~\cite{Cichy2019}.
\begin{figure}[h!]
    \center
    \includegraphics[width=12cm,height=5cm,keepaspectratio]{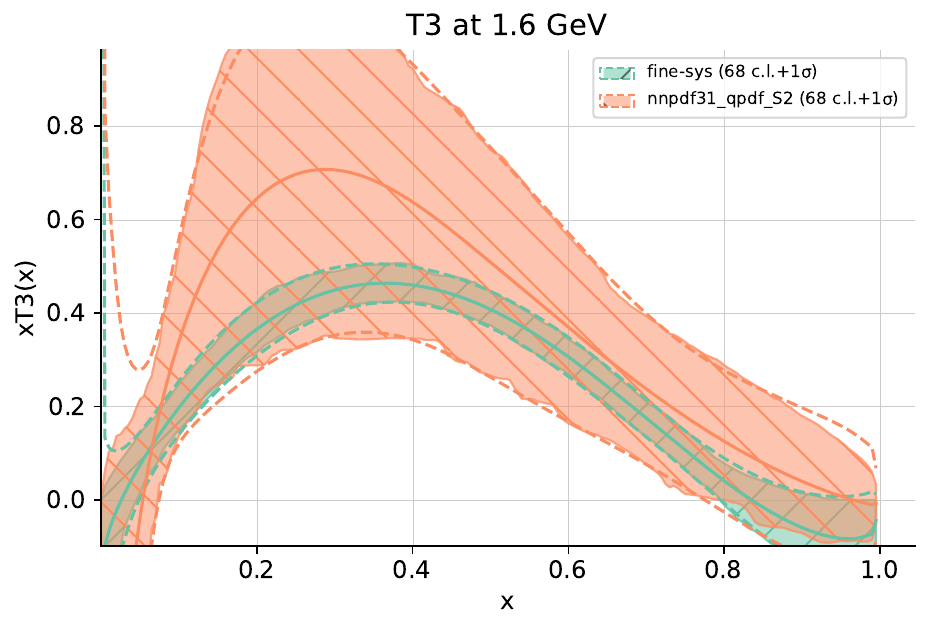}
    \includegraphics[width=12cm,height=5cm,keepaspectratio]{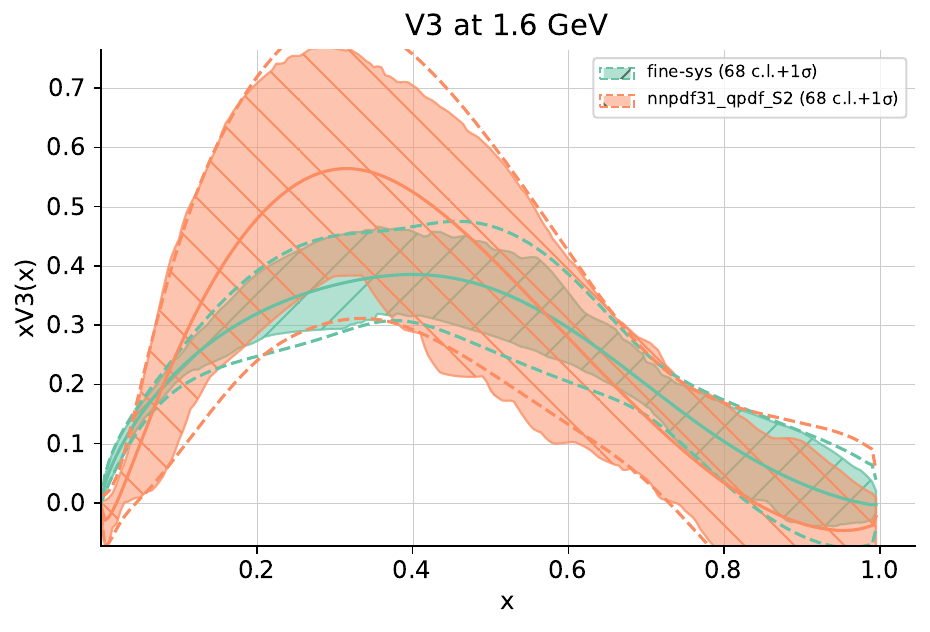}
    \caption{PDFs from the fits \textit{fine-sys} compared with the corresponding distributions from 
    the fit nnpdf31\_qpdf\_S2, presented in Ref.~\cite{Cichy2019}.}
    \label{fig::ppdf_vs_qpdf}
\end{figure} 
Quasi-PDFs and pseudo-ITD results are plotted together in Fig.~\ref{fig::ppdf_vs_qpdf}: 
both $T_3$ and $V_3$ distributions appear to be in good agreement, the main difference being a huge decrease in the 
PDFs error when considering results presented in this work. 
This difference can be partially traced back to the number of points included in the analysis: while in Ref.~\cite{Cichy2019}
16 points for quasi-PDFs matrix element where included, in the present work data corresponding to all momentum values are considered,
for a total of 48 pseudo-ITD points.
Clearly, having more points in the analysis allows to better constraint the fit results, giving final PDFs with smaller error.
 Given equivalent computational cost, the low momenta matrix elements, which are used in the pseudo-PDF approach, are exponentially more precise than the large momenta matrix elements, to which the quasi-PDF approach are restricted. The size of the statistical and systematic uncertainties affecting the points entering the two analyses is of course another
reason for different PDFs error, however a detailed study of such differences is beyond the scope of this work.
We leave a detailed comparison between the quasi- and pseudo-PDFs approaches for a future study.